\documentclass[12pt]{cernart}   
\usepackage{epsfig}
\usepackage{here}
\usepackage{amssymb}
\usepackage{graphicx}

\begin{document}

%%%%%%%%%%%%%%%%%%%%%%%%%%%%%%% titlepage %%%%%%%%%%%%%%%%%%%%%%%%%%%%%%%%%%%%
\begin{titlepage}
%  \docnum{{\tt NA48 Note 99--25}}
  \docnum{CERN--EP/99--139}
  \date{7 October 1999}
  \title{Sceptical combination of experimental results:\\ 
General considerations and
application to
$\epsilon^\prime/\epsilon$ }
        \author{G. D'Agostini\Instref{xx}}
% Submitted{}
% collaboration{}
% conference{}
% note{}
% dedication{}
\Instfoot{xx}{ Universit{\`a} `La Sapienza' and
        Sezione INFN di Roma 1, Rome, Italy, and CERN, Geneva, Switzerland\\
        {\rm Email}: {\tt dagostini@roma1.infn.it}\\ 
        {\rm URL}: {\tt http://www-zeus.roma1.infn.it/$^\sim$agostini}}
\begin{abstract}
This paper is meant as a contribution to the often debated 
subject of how to 
combine data which appear to be in mutual 
disagreement. As a practical example, the 
$\epsilon^\prime/\epsilon$ determinations have been considered. 
\end{abstract}
\vspace*{12.5cm}
\begin{center}
{\it (Submitted to Physical Review D)}
\end{center}
\end{titlepage}

\section{Introduction}
\setcounter{page}{2}
Every physicist  knows the 
rule for combining several experimental results:
\begin{eqnarray}
\mu &=& \frac{\sum_id_i/s_i^2}{\sum_i1/s_i^2} 
\label{eq:media}\\
\sigma(\mu) &=& \left(\sum_i1/s_i^2\right)^{-\frac{1}{2}}\,,
\label{eq:sigma}
\end{eqnarray}
where `$\mu$' refers to the true value and
$d_i \pm s_i$ stands for the individual data 
point (the use of $s_i$, instead of the usual $\sigma_i$, 
for the standard uncertainty reported by the experiments will become 
clear later; similarly, the meaning of `$\mu$' and of $\sigma(\mu)$ 
have  not been well defined for the moment, as they will be better defined
later). 
The above rule, hereafter called {\it standard combination rule}, 
is based on some hypotheses which are worth recalling:\break\hfill
{\it i)} all measurements refer to the same quantity;
{\it ii)}  the measurements are independent;
{\it iii)} the probability distribution of $d_i$ around $\mu$ is described 
by a Gaussian distribution with standard deviation 
given by $\sigma_i=s_i$.
If one, or several, of these 
 hypotheses are not satisfied, 
the result of  Eqs.~(\ref{eq:media})--(\ref{eq:sigma}) is questionable.
 
Now we are confronted with the problem
that we are never absolutely sure 
if these hypotheses are true or not. If we were absolutely convinced 
that the hypotheses were correct, there would be no reason to hesitate 
to apply Eqs. (\ref{eq:media})--(\ref{eq:sigma}), no matter 
how `apparently incompatible' the data points might appear. 
But we know by experience
that unrecognized sources of systematic errors
might affect the results, or that the uncertainty associated with the 
recognized sources might be underestimated (but we also know that, 
often, this kind of uncertainty is prudently overstated\ldots). 

As   is always the case in the domain of   uncertainty, 
there is no `objective' method for   handling this problem; 
  neither in deciding if 
the data are in mutual disagreement, nor in 
arriving at a universal solution for handling those
cases which are judged to be troublesome.
Only good sense gained by experience can 
provide   some guidance. Therefore, 
all automatic `prescriptions'
should be considered {\it cum grano salis}. 
For example, the usual method for 
checking the hypothesis that `the data are compatible with each other' is to
make a $\chi^2$ test. The hypothesis is accepted if, generally
speaking, the $\chi^2$ does not differ too much from the 
expected value. 
As a strict rule, the $\chi^2$ test is not really logically grounded  
(see e.g. Section 1.8 of Ref.~\cite{YR}) although it does `often work',
due to implicit hypotheses which are 
external to the standard $\chi^2$ test  
scheme (see Section 8.7 of Ref.~\cite{YR}), but which 
lead to mistaken conclusions when 
the unstated hypotheses are not reasonable
(see e.g. Section 1.9 of Ref.~\cite{YR}). 
Therefore, I shall not attempt here to quantify the 
degree of suspicion. I shall assume  a situation 
in which experienced physicists, faced with a set of results,
tend to be uneasy about the mutual consistency of the picture
that those data offer. 

\begin{figure}
\begin{center}
\begin{tabular}{|c|}\hline
\epsfig{file=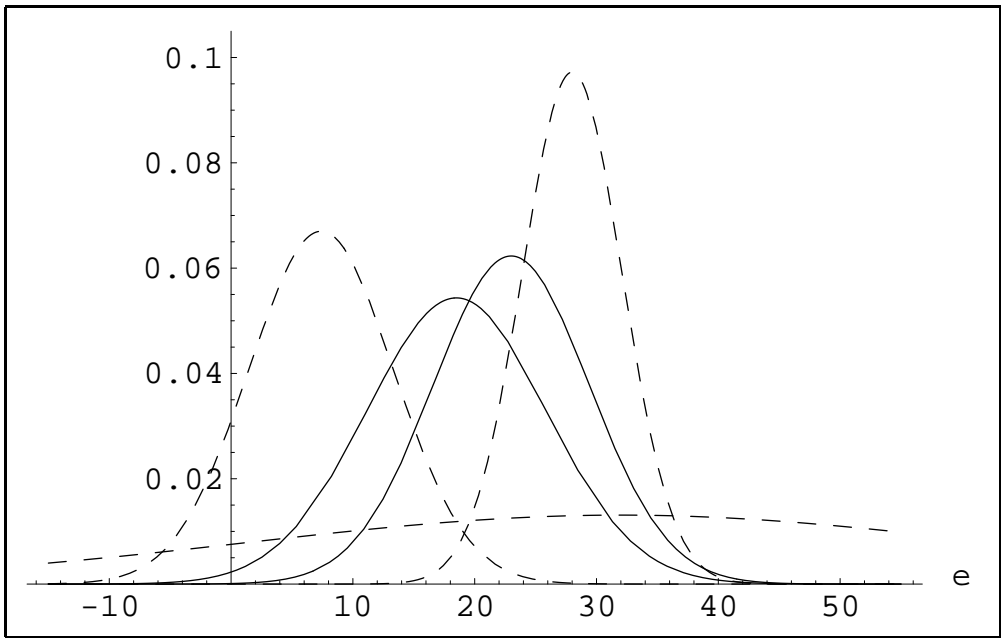,width=0.8\linewidth,clip=} 
\\ \hline
\end{tabular}
\end{center}
\caption{\small Results on Re($\epsilon^\prime/\epsilon$)
obtained at CERN (solid line) and Fermilab (dashed line), where 
$e={\rm Re}(\epsilon^\prime/\epsilon)\times 10^{4}$.}
\label{fig:results}
\end{figure}

\begin{figure}[t]
\begin{center}
\begin{tabular}{|c|}\hline
\epsfig{file=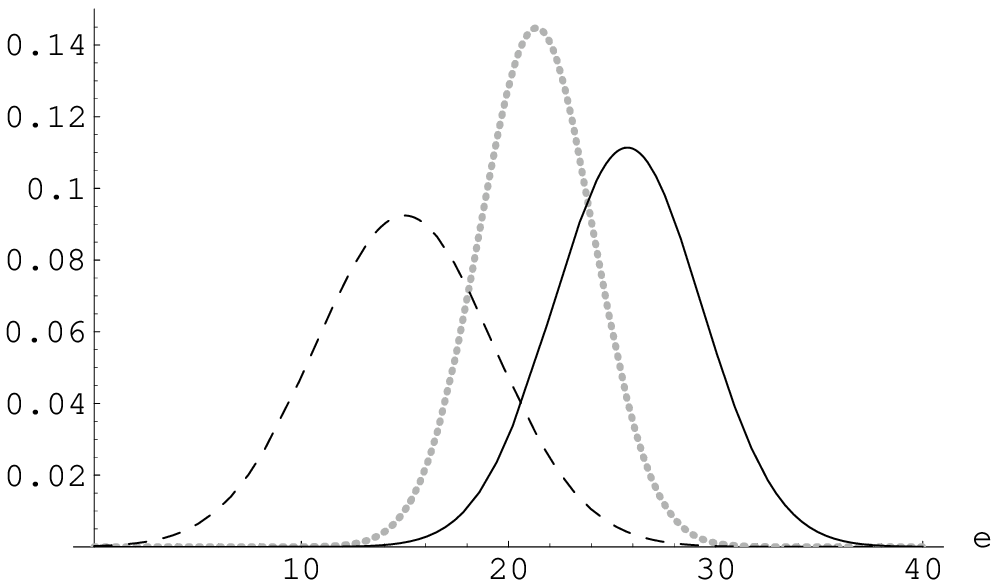,width=0.63\linewidth,clip=}  \\ 
\hline
\epsfig{file=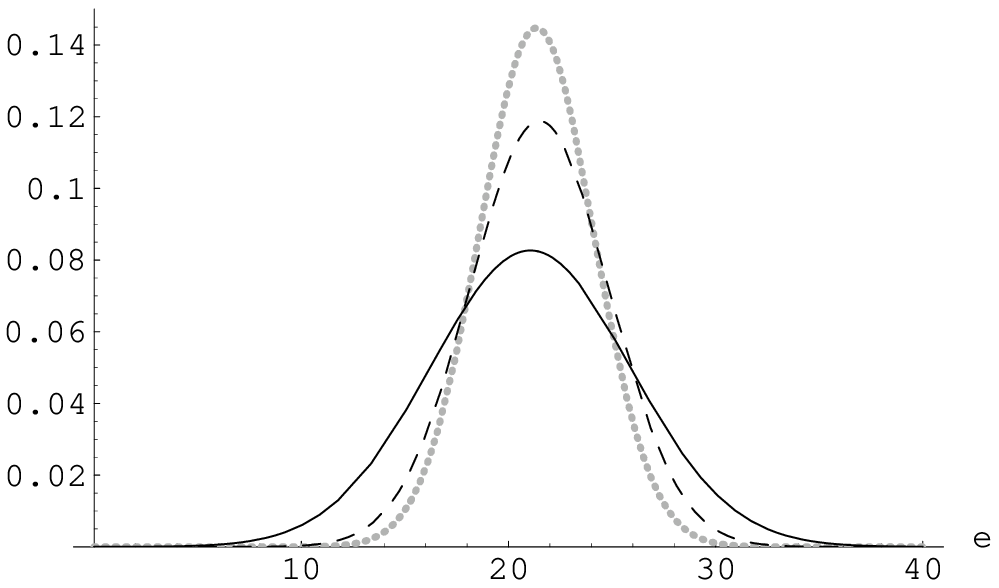,width=0.63\linewidth,clip=} \\ \hline
\end{tabular}
\end{center}
\caption{\small Some combinations of the experimental results 
obtained using the
standard combination rule 
of  Eqs.~(\ref{eq:media})--(\ref{eq:sigma}). 
Upper plot:  old results (dashed line), 1999 results (solid line), 
overall combination (dotted grey line). Lower plot: CERN experiments 
(solid line), Fermilab experiments (dashed), overall combination 
(dotted grey line).} 
\label{fig:naive}
\end{figure}

As an example, let us consider 
the results of Table~\ref{tab:results},
\begin{table}
\caption{\small Published results on 
Re($\epsilon^\prime/\epsilon$) (values in units of $10^{-4}$). 
Data points indicated by $\surd$ have been used for quantitative 
evaluations.
Owing to correlations between the 1988 and 1993 uncertainties 
of NA31, only 
the combined value published in 1993 is used protect{\cite{Wahl}}.}
\label{tab:results}
\begin{center}
\begin{tabular}{|ll|cc|c|}\hline
&&&& \\
\multicolumn{2}{|c|}{Experiment} & 
Central value          & $\pm\sigma_{\mbox{\it stat}}
\pm\sigma_{\mbox{\it syst}}$       &  $\sigma_{\mbox{\it tot}}$ \\
&&&& \\ \hline
&&&& \\
$\surd$ & E731 (1988) \cite{E73188}          & 32   & $ \pm 28 \pm 12$  & 30 \\
        & NA31 (1988)\cite{NA3188}           & 33   & $\pm6.6\pm8.3$    & 11 \\
$\surd$ &E731 (1993)\cite{E73193}           & 7.4  & $\pm5.2\pm2.9$    & 5.9 \\
        & NA31 (1993)\cite{NA3193}           & 20   & $\pm4.3\pm5.0$    & 7 \\
$\surd$ &NA31 (1988+1993)\cite{NA3193,Wahl} &23.0  &  $\pm4\pm5$     & 6.5  \\
$\surd$ &KTeV (1999)\cite{KTeV}            & 28.0 & $\pm3.0\pm2.8$    & 4.1  \\
$\surd$ &NA48 (1999)\cite{NA48}       & 18.5 & $\pm4.5\pm5.8$    & 7.3  \\
&&&& \\
\hline
\end{tabular}
\end{center}
\end{table}
 which are also
reported in a graphical form in Fig. \ref{fig:results}. 
Figure \ref{fig:naive} shows also the combined result 
obtained using\break\hfill  Eqs. (\ref{eq:media})--(\ref{eq:sigma}),
as well as some combinations of subsamples of the results. 
These results have not been chosen as the best example
of disagreeing data, but because of the physics interest, and
also because the situation is  at the edge of where 
one starts worrying.  The impression of uneasiness is not only  
because   the mutual agreement among the experimental results
is not at the level one would have wished, but also 
because the value of Re($\epsilon^\prime/\epsilon$) around which 
the experimental results cluster is somewhat far from the
theoretical 
evaluations (see e.g. Refs.~\cite{PhysW,Buras,Fabbrichesi,Roma99,Nierste} 
and references therein).  
Now, it is clear that experimentalists should not be biased towards 
theoretical expectations, and the history of physics teaches us about 
wrong results published to please theory. But we are 
also aware of unexpected results (either claims of 
new  physics, or simply a quantitative disagreement
with respect to the global scenario offered by other results 
within the framework of the Standard Model)
which finally turn out to be false alarms. In conclusion, 
given the present picture of theory
versus experiments about $\epsilon^\prime/\epsilon$, there is 
plenty of room for doubt: Doubt about theory, about individual experiments, 
and about experiments as a whole. 

In this situation, drawing conclusions based on a blind application 
of Eqs. (\ref{eq:media})--(\ref{eq:sigma}) seems a bit na\"\i ve. 
For example, a straightforward conclusion of the standard 
combination rule 
leads to a probability that 
Re($\epsilon^\prime/\epsilon$) is smaller than zero of the order of
$0.5~\times~10^{-14}$, and I don't think that experienced physicists would
share  without hesitation beliefs of this order of magnitude.  

This paper deals with modelling the beliefs of an
experienced  {\it sceptical physicist} 
confronted with results of this kind, continuing on from 
a recent work 
of Dose and von der Linden on {\it outliers}~\cite{Dose}. 

\section{Hypotheses behind the simple combination rule}
Equation (\ref{eq:media}) has been written, on purpose, 
in a way that  might be misleading, although 
this is the way in which it  often appears.
In fact, taken literally, 
it says that $\mu$ {\it is} equal to the right-hand side of 
Eq.~(\ref{eq:media}). Instead, as is well understood, 
this is just the value around 
which our beliefs are centred, usually referred to as  the {\it estimator}. 
Given a Gaussian model, 
the estimator given by Eq.~(\ref{eq:media})
 corresponds to the value which we believe mostly ({\it mode}), 
and also to the barycentre of the probability 
distribution\footnote{Following   physics intuition, 
we consider it natural to talk about 
probability of true values. 
For historical reasons, this point of view is currently 
known by the somewhat esoteric name of Bayesian, to distinguish it 
from the so-called frequentistic point of view, according to
which the category of probable should not be applied
to true values and, generally speaking, to 
hypotheses. For a physicist's introduction to Bayesian reasoning  
see Ref.~\cite{YR}, or Ref.~\cite{ajp} for a short account.} of $\mu$ 
({\it expected value}) and to the value which defines two
semi-open intervals in each of which we believe $\mu$ to lie  with 
equal probability ({\it median}). 

In order to obtain a combination rule different from 
Eqs. (\ref{eq:media})--(\ref{eq:sigma}), it is important 
to remember where these formulae  come from. Although this rule 
is usually taught in the framework of maximum likelihood, the most general 
way to get it is by using Bayesian inference, as  we shall 
show now.
 
The simplest way to write Bayes' theorem for continuous 
variables is:  
\begin{equation}
f(\mu\,|\,\underline{d}) \propto f(\underline{d}\,|\,\mu)\cdot f_\circ(\mu)\,,
\label{eq:Bayes}
\end{equation}
where the set of data points 
$\{d_1, d_2,\ldots, d_n\}$ is indicated by
$\underline{d}$; the  function  
$f(\mu\,|\,\underline{d})$   is the {\it final} 
probability density function (p.d.f.) 
  of $\mu$ in the light of the experimental results 
and of all other prior knowledge about measurement and measurand; 
$f(\underline{d}\,|\,\mu)$ represents the {\it likelihood} 
of observing the data set $\underline{d}$ under the hypothesis 
that the true value is exactly $\mu$; $f_\circ(\mu)$ 
is the {\it prior} p.d.f. of $\mu$. The proportionality factor
is obtained  by the normalization condition 
$\int\!f(\mu\,|\,\underline{d})\,\mbox{d}\mu=1$.  
The assumption that each of the observed values is normally
distributed around $\mu$ with standard deviation $\sigma_i$
and that the measurements are 
independent leads to 
\begin{equation}
f(\underline{d}\,|\,\mu) = \prod_i\frac{1}{\sqrt{2\,\pi}\,\sigma_i}
\,\exp\left[-\frac{(d_i-\mu)^2}{2\,\sigma_i^2}\right]\,.
\label{eq:likelihood}
\end{equation} 
If the experimental resolution
described by the likelihood is sufficiently high
and $\mu$ is a quantity which can assume, 
in principle, values in a large interval (virtually any real values), 
a uniform prior distribution, i.e. $f_\circ(\mu) = k$,
 is a very reasonable assumption.
In fact, any other mathematical function which models 
the vagueness of the prior knowledge (with respect to what 
the measurement is supposed to yield) acts in practice as 
a constant in the region of $\mu$ where the likelihood 
varies rapidly.  
Putting all the ingredients together 
and renormalizing the final p.d.f.
we get
\begin{equation}
f(\mu\,|\,\underline{d},\mbox{indep. Gaussians},
\underline{\sigma}, f_\circ(\mu)=k) = 
\frac{1}{\sqrt{2\,\pi}\,\sigma(\mu)}\,
\exp\left[-\frac{(\mu-\mbox{E}[\mu])^2}{2\,\sigma^2(\mu)}\right]\,,
\label{eq:gauss_comb}
\end{equation}
where 
\begin{eqnarray}
\mbox{E}[\mu] &=& \frac{\sum_id_i/s_i^2}{\sum_i1/s_i^2} 
\label{eq:media_E}\\
\sigma(\mu) &=& \left(\sum_i1/s_i^2\right)^{-\frac{1}{2}}\,,
\end{eqnarray}
obtained assuming that the $\sigma_i$ of 
Eq.~(\ref{eq:likelihood}) are exactly equal to the quoted stated
uncertainties $s_i$.
In Eq.~(\ref{eq:gauss_comb}) all conditions have been 
explicitly stated. 
This derivation shows that there is indeed a fourth  important 
implicit assumption in order to arrive at 
Eqs. (\ref{eq:media})--(\ref{eq:sigma}), 
namely a uniform prior\footnote{For those used  to frequentistic
methods, in which `there are no priors', I would like 
to recall how Gauss \cite{Gauss} 
derived his famous Gaussian distribution
describing experimental errors.
He made explicit use of the concepts of prior 
and posterior probability of hypotheses,
and derived a formula equivalent to   Bayes' theorem valid 
for a priori equiprobable hypotheses (condition 
explicitly stated). 
Then, using some symmetry arguments, plus the condition that the final
distribution is maximized when the true value of the quantity equals the
arithmetic average of the measurements, he obtained 
the functional form of the
error distribution (playing the role of likelihood), which is now named after
him.}
 on $\mu$. 
This is   why the maximum belief coincides
with the maximum of the likelihood, and why the 
best estimate of $\mu$ is the same as is obtainable  
from the maximum likelihood principle. Nevertheless, the 
route followed here is more general and more intuitive, as discussed 
extensively in Ref.~\cite{YR}. In particular, one can speak 
consistently about probability of true values, a concept close
to the natural reasoning of physicists~\cite{maxent98}.

\section{Probabilistic modelling of scepticism}
Once we have understood what is behind the simple combination
rule, it is possible to change one of the hypotheses entering 
Eq.~(\ref{eq:gauss_comb}). Obviously, the problem 
has no unique solution. This 
depends to a great extent on the status of knowledge about the experiments
which provided the results. For example, if one has
formed a personal idea concerning
the degree of reliability of the different experimental teams,
one can attribute different weights 
to different results, or even disregard results   
considered unreliable or obsolete
(for example their corrections for systematic effects 
could depend on theoretical inputs
which are now considered to be obsolete). 
Wishing to arrive at a solution which, with all
the imaginable limitations a general solution may have, 
is applicable to many situations
without an inside, detailed knowledge of each individual 
experiment, we have to make some 
choices. First, we decide that {\it our sceptic is democratic},  
i.e. `he' has no a priori
preference for a particular experiment.  
Second, the easiest way of modelling his
scepticism, keeping the  mathematics simple, 
is to consider the likelihood still
Gaussian, but with a standard deviation
which might differ from that quoted by the experimentalists
by a factor $r_i$ which is not exactly known:
\begin{equation}
r_i = \frac{\sigma_i}{s_i}~.
\end{equation}
The uncertainty about $r_i$ can be described by a p.d.f. $f(r_i)$. 
This uncertainty changes each factor appearing 
in the likelihood (\ref{eq:likelihood}), 
as  can be evaluated by the probability rules: 
\begin{equation}
f(d_i\,|\,\mu) = 
\int\!f(d_i\,|\,\mu,r_i)\cdot f(r_i)\,\mbox{d}r_i\,,
\label{eq:lik_i}
\end{equation}
with 
\begin{equation}
f(d_i\,|\,\mu,r_i,s_i) = \frac{1}{\sqrt{2\,\pi}\,r_i\,s_i}\,
          \exp\left[-\frac{(d_i-\mu)^2}{2\,r_i^2\,s_i^2}\right]\,.
\label{eq:lik_i_r}
\end{equation}
If one believes that all $r_i$ are exactly one, i.e. 
$f(r_i)=\delta(r_i-1)\ \forall\, i$,
the standard combination rule is recovered. 
Because of our basic assumption of democracy,
the mathematical expression of the p.d.f. of $r_i$ 
will not depend on $i$, therefore we shall talk hereafter,
generically, about $r$ and $f(r)$.   

A solution to the problem of finding
a parametrization of $f(r)$ such that this p.d.f. is acceptable 
to experienced physicists, even though the integral (\ref{eq:lik_i})
still has a closed form, 
has been proposed by Dose and von der Linden~\cite{Dose}; 
an improved version of it will be used in this paper~\cite{Dose1}. 
Following Ref.~\cite{Dose}, we choose initially
the variable $\omega=1/r^2=s_i^2/\sigma_i^2$, 
and  consider it to be described by a 
gamma distribution: 
\begin{equation}
f(\omega) = 
\frac{\lambda^\delta\,
\omega^{\delta-1}\,e^{-\lambda\,\omega}}{\Gamma(\delta)}\,,
\end{equation}
where $\lambda$ and $\delta$ are the so-called 
 scale and shape parameters, respectively. 
As a function of these two parameters, expected value and 
variance of $\omega$ 
are $\mbox{E}[\omega]=\delta/\lambda$ and 
$\mbox{Var}(\omega)=\delta/\lambda^2$.
Using probability calculus we get the p.d.f of $r$:  
\begin{equation}
f(r\,|\,\lambda,\delta) = 
\frac{2\,\lambda^\delta\,
r^{-(2\,\delta+1)}\,e^{-\lambda/r^2}}{\Gamma(\delta)}\,,
\label{eq:r_Dose1}
\end{equation}
where the parameters have been written explicitly as 
conditionands for the probability distribution. Expected value
and variance of $r$ are: 
\begin{eqnarray}
\mbox{E}[r]   &=& \frac{\sqrt{\lambda}\,\Gamma(\delta-1/2)}
                     {\Gamma(\delta)}\\
\mbox{Var}(r) &=& \frac{\lambda}{\delta-1} 
 -\frac{\lambda\,\Gamma^2(\delta-1/2)}
                     {\Gamma^2(\delta)}\,,
\end{eqnarray}
existing simultaneously if $\lambda > 0$ and $\delta>1$. 

The individual likelihood, integrated over the possible values of $r$,
is obtained  by inserting Eqs. (\ref{eq:lik_i_r}) and 
(\ref{eq:r_Dose1}) 
 in Eq.~(\ref{eq:lik_i}): 
\begin{equation}
f(d_i\,|\,\mu,s_i) = \frac{\lambda^\delta}{\sqrt{2\,\pi}s_i}\,
                 \frac{\Gamma(\delta+1/2)}{\Gamma(\delta)}\,
       \left(\lambda+\frac{(d_i-\mu)^2}
                          {2\,s_i^2}\right)^{-(\delta+1/2)}\,. 
\label{eq:int_lik2}
\end{equation}
Using a uniform prior distribution for $\mu$,
and remembering that we are dealing with 
independent results, we have finally: 
\begin{equation}
f(\mu\,|\,\underline{d},\underline{s})\propto 
f(\underline{d}\,|\,\underline{s},\mu) 
\propto \prod_i 
\left(\lambda+\frac{(d_i-\mu)^2}{2\,s_i^2}\right)^{-(\delta+1/2)}\,,
\label{eq:final_DL2}
\end{equation}
where $\underline{s}=\{s_1, s_2,\ldots, s_n\}$. 
The normalization factor can be determined numerically.
Equation~(\ref{eq:final_DL2})
should be written, more properly, as 
$f(\mu\,|\,\underline{d}, \underline{s},\lambda, \delta)$, 
to remind us that the solution depends 
on the choice of $\lambda$ and $\delta$,
and teaches us how to get a solution which takes into account
all reasonable choices of the parameters: 
\begin{equation}
f(\mu\,|\,\underline{d}, \underline{s}) = 
\int\!f(\mu\,|\,\underline{d}, \underline{s},
\lambda, \delta)\cdot f(\lambda, \delta)
\,\mbox{d}\lambda\mbox{d}\delta\,,
\label{eq:intld}
\end{equation}
where $f(\lambda, \delta)$ quantifies the confidence
on each possible pair of parameters.\footnote{$\lambda$ 
and $\delta$ are the same for all experiments as we are 
modelling a democratic scepticism. In general they could 
depend on the experiment, thus changing  
Eq.~(\ref{eq:final_DL2}).}

\begin{figure}
\begin{center}
\begin{tabular}{|c|}\hline
\epsfig{file=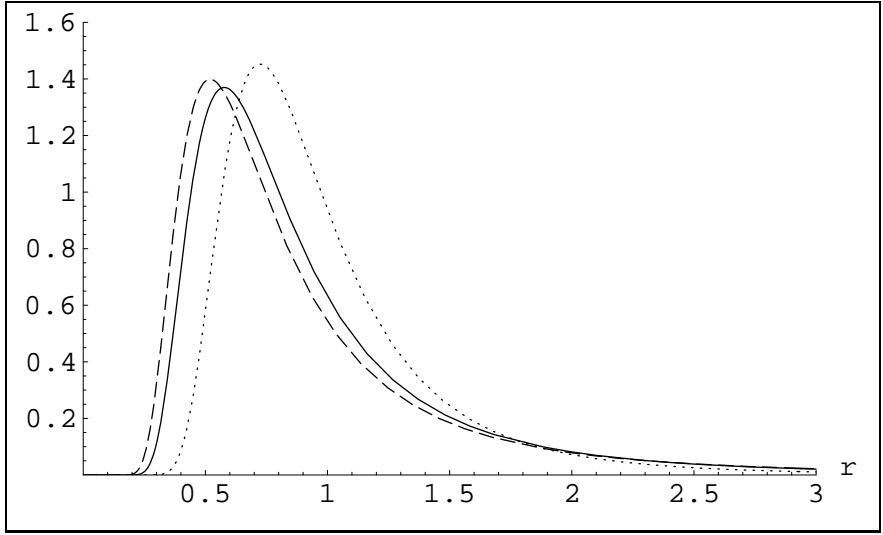,width=0.7\linewidth,clip=} \\ \hline
\end{tabular}
\end{center}
\caption{\small Distribution of the rescaling factor 
$r=\sigma_{\mbox{true}}/\sigma_{\mbox{est}}$
using the parametrizations of Eq.~(\ref{eq:r_Dose1})
for several values of the set of parameters $(\lambda,\delta)$; 
the solid line 
corresponds to what will be taken as the reference distribution 
in this paper, yielding $\mbox{E}[r]=\sigma(r)=1$, 
and it is obtained for $\lambda\approx 0.6$ and $\delta\approx 1.3$. 
Dotted and dashed lines  show the p.d.f.
of $r$ yielding $\sigma(r)=0.5$ and 1.5, respectively.} 
\label{fig:rDL2}
\end{figure}

\begin{figure}
\begin{center}
\begin{tabular}{|c|c|}\hline
 Individual results  &  Eq.~(\ref{eq:final_DL2}), 
$\mathbf{\lambda=0.6}$ and $\mathbf{\delta=1.3}$ \\
      & $[\mathbf{\sigma(r) = 1}]$ \\ \hline 
\epsfig{file=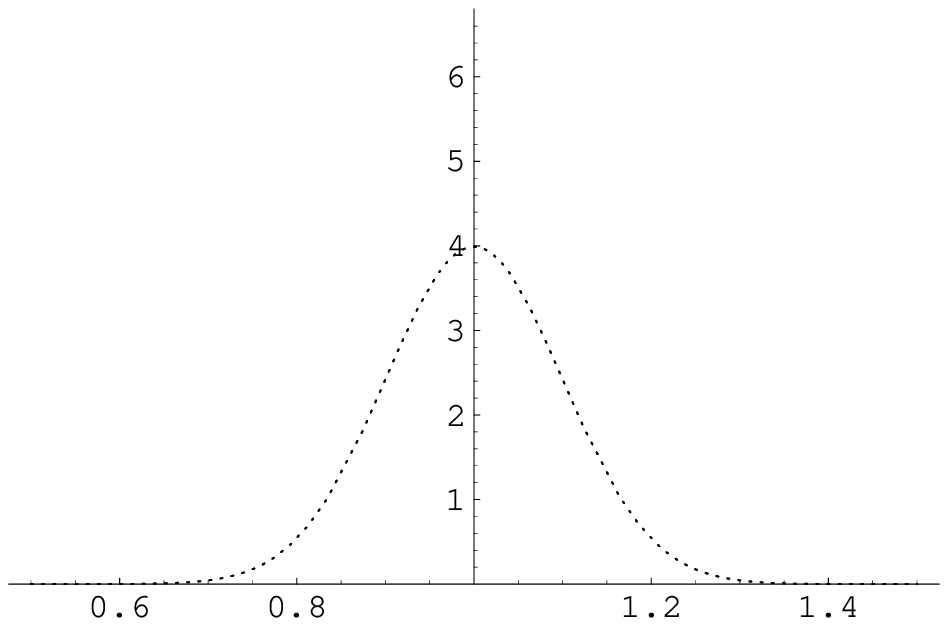,width=0.4\linewidth,clip=}  &
\epsfig{file=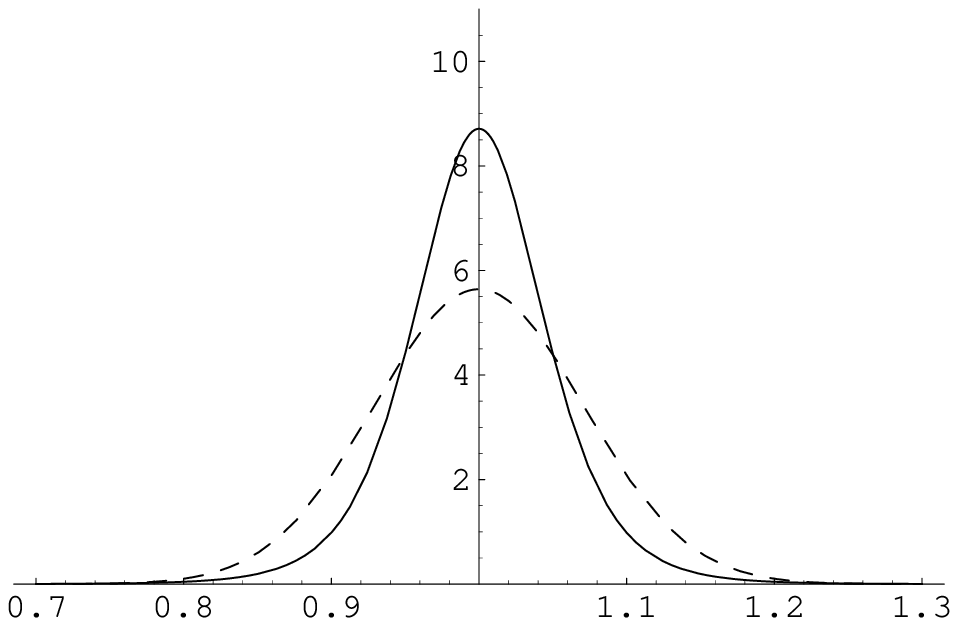,width=0.4\linewidth,clip=}  \\\hline
\epsfig{file=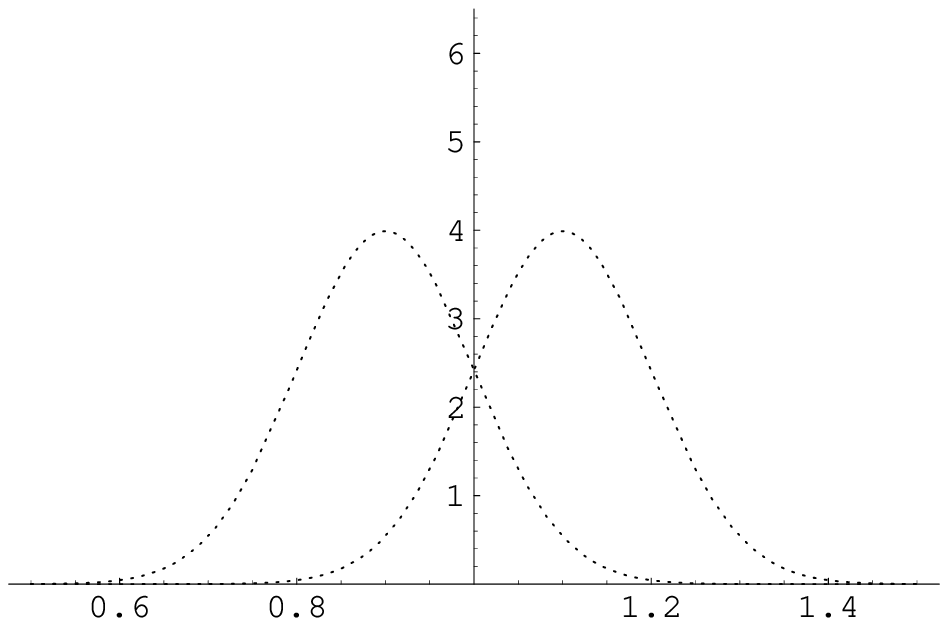,width=0.4\linewidth,clip=}  &
\epsfig{file=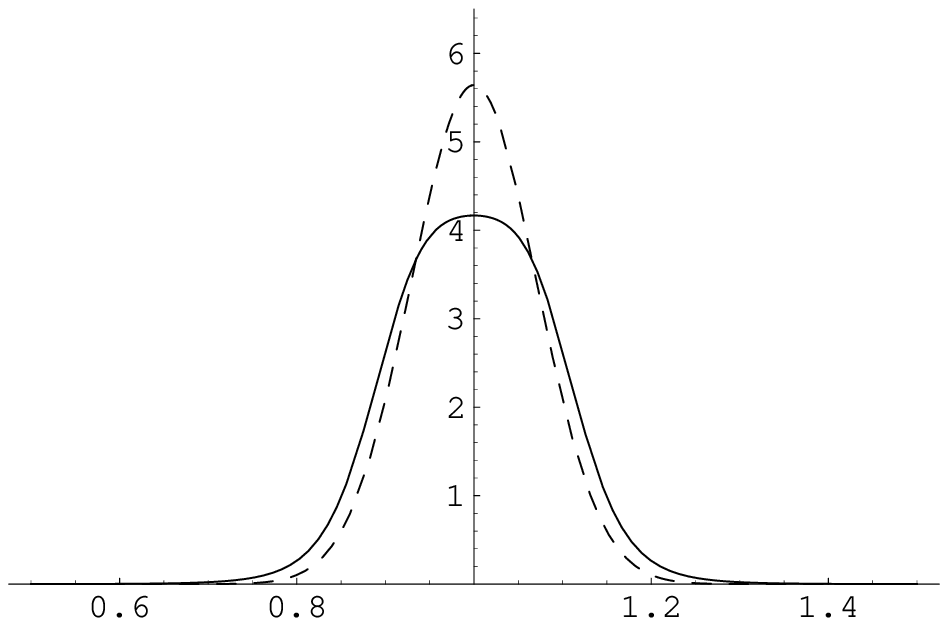,width=0.4\linewidth,clip=}  \\\hline
\epsfig{file=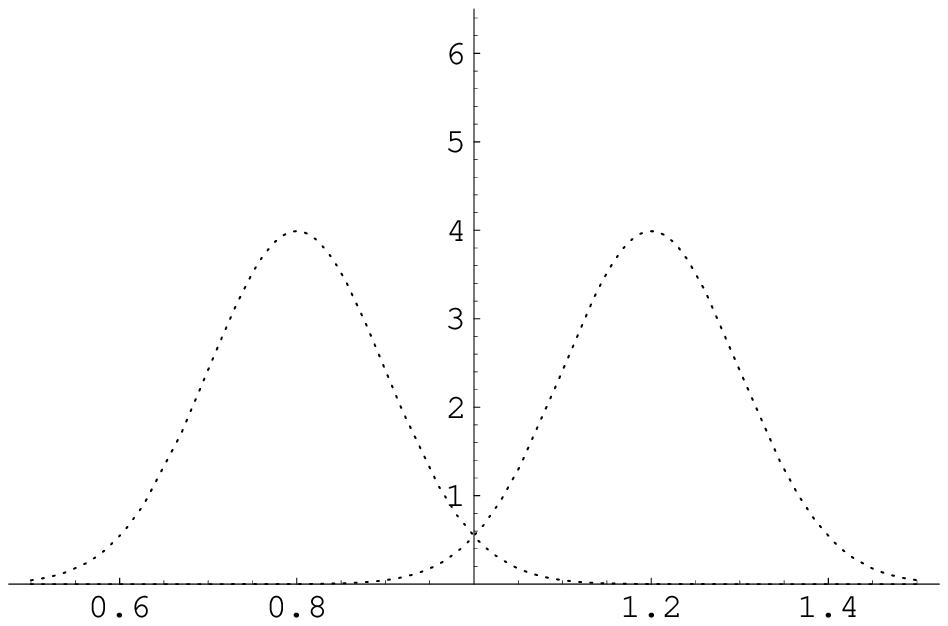,width=0.4\linewidth,clip=}  &
\epsfig{file=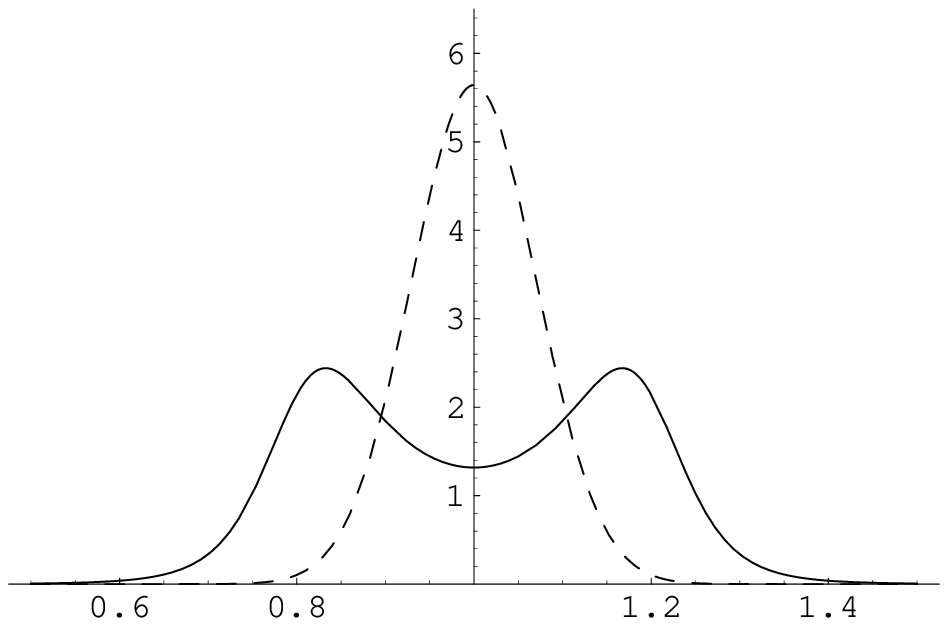,width=0.4\linewidth,clip=}  \\\hline
\epsfig{file=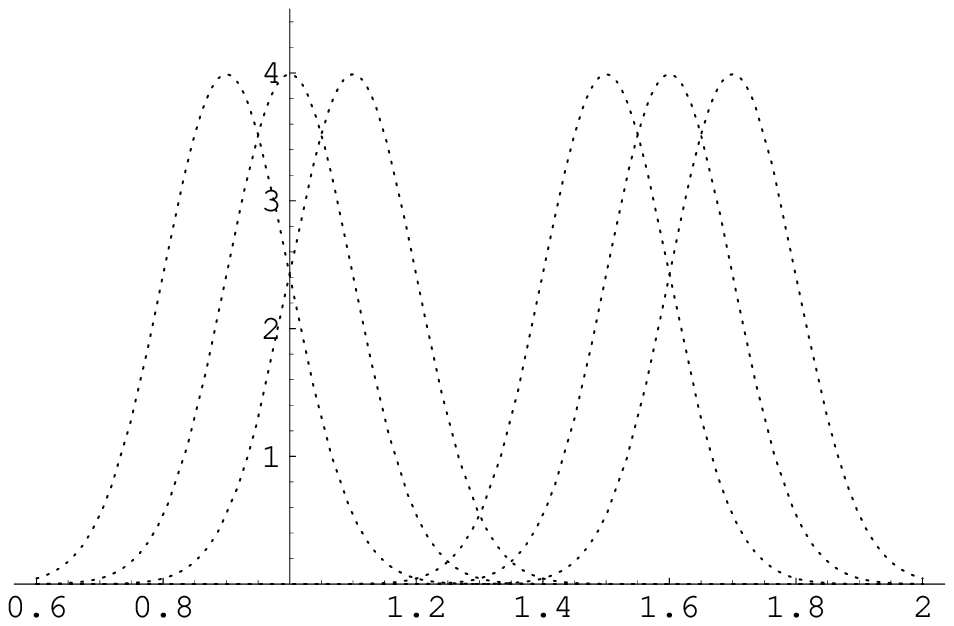,width=0.4\linewidth,clip=}  &
\epsfig{file=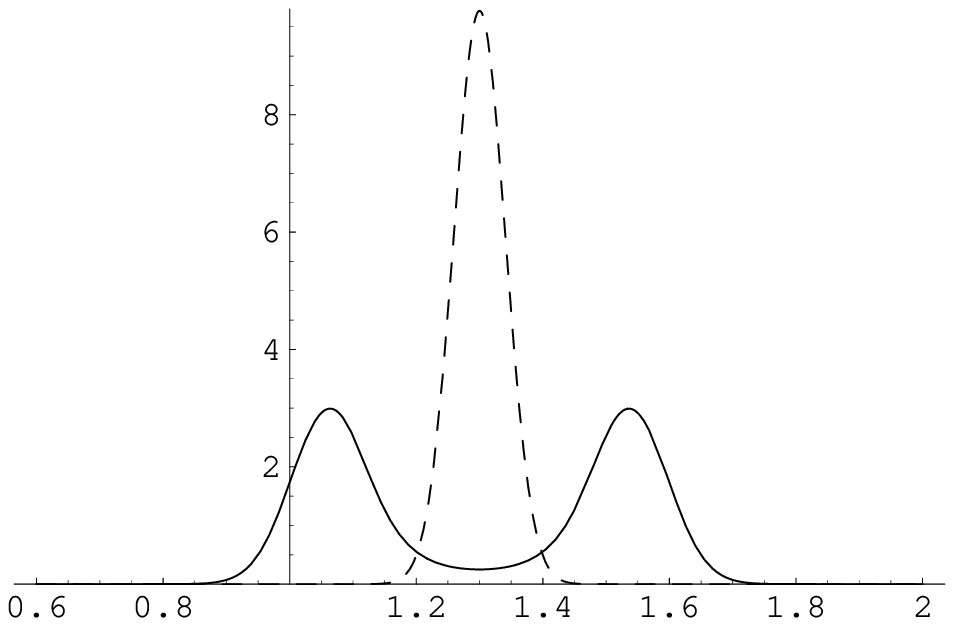,width=0.4\linewidth,clip=} \\ \hline
\epsfig{file=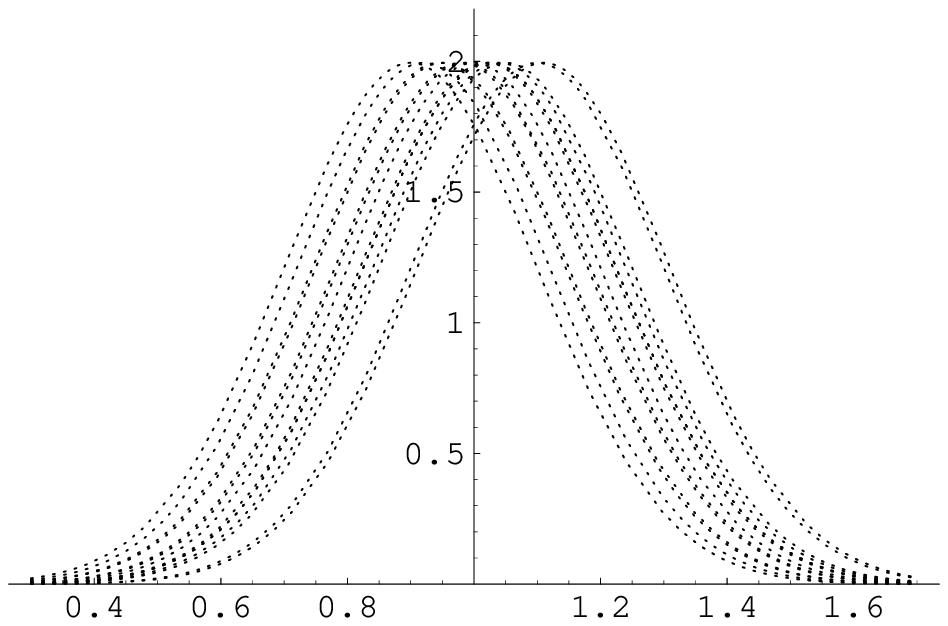,width=0.4\linewidth,clip=}  &
\epsfig{file=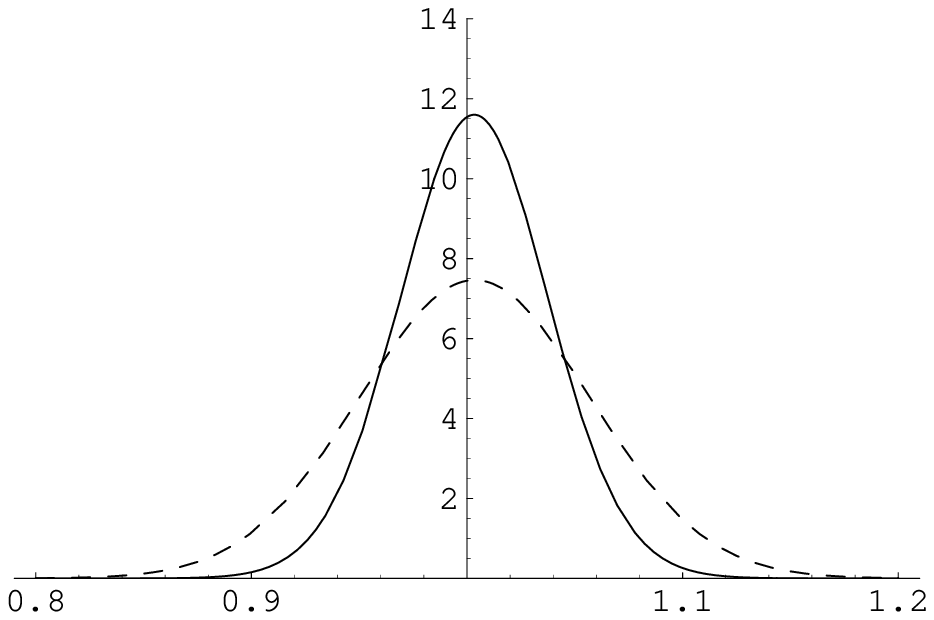,width=0.4\linewidth,clip=} \\ \hline
\end{tabular}
\end{center}
\vspace{-18.5cm}
\mbox{}\hspace{2.5cm}
{\Large $2\,\times$}
\vspace{+17.9cm}
\caption{\small Examples of sceptical combination of results. 
The plots on the left-hand side  show the individual results (in the upper
plot the two results coincide).
The  plots on the right-hand side  show the combined result obtained 
using  Eq.~(\ref{eq:final_DL2}) with the constraint 
$\mbox{E}[r]=\sigma(r)=1$ (continuous lines), 
compared with the standard combination (dashed lines).} 
\label{fig:discordant}
\end{figure}

A natural constraint on the values of the parameters
comes from the request\break\hfill $\mbox{E}[r]=1$, modelling the
assumption that the $\sigma$'s agree, on average, with 
the stated\break\hfill\newpage\noindent uncertainties. 
The standard deviation of the distribution gives 
another constraint. Conservative considerations 
suggest $\sigma(r)/\mbox{E}[r]\approx {\cal O}(1)$. 
The condition $\mbox{E}[r] = \sigma(r)$ = 1
is obtained for $\lambda \approx 0.6$ and $\delta \approx 1.3$. 
The resulting p.d.f. of
$r$ is shown as the continuous line of Fig.~\ref{fig:rDL2}. 
One can see that the parametrization of $f(r)$ corresponds 
qualitatively to intuition: the barycentre of the distribution 
is 1; values below $r\approx 1/2$ are considered practically impossible;
on the other hand, very large values of $r$ are conceivable, although
with very small probability, indicating that 
large overlooked systematic errors might occur.   
Anyway, we feel that, 
besides general arguments and considerations about  
the shape of $f(r)$ (to which we are not used),  
what matters is how reasonable the 
results look. Therefore, the method has been tested 
with simulated data, shown in the left plots of  
Fig.~\ref{fig:discordant}.

\begin{figure}
\begin{center}
\begin{tabular}{|c|c|}\hline
Eq.~(\ref{eq:final_DL2}), $\lambda=1.4$ and $\delta=2.1$ &
Eq.~(\ref{eq:final_DL2}), $\lambda=0.4$ and $\delta=1.1$ \\
$[\sigma(r) = 0.5]$ & $[\sigma(r) = 1.5]$ \\ \hline
\epsfig{file=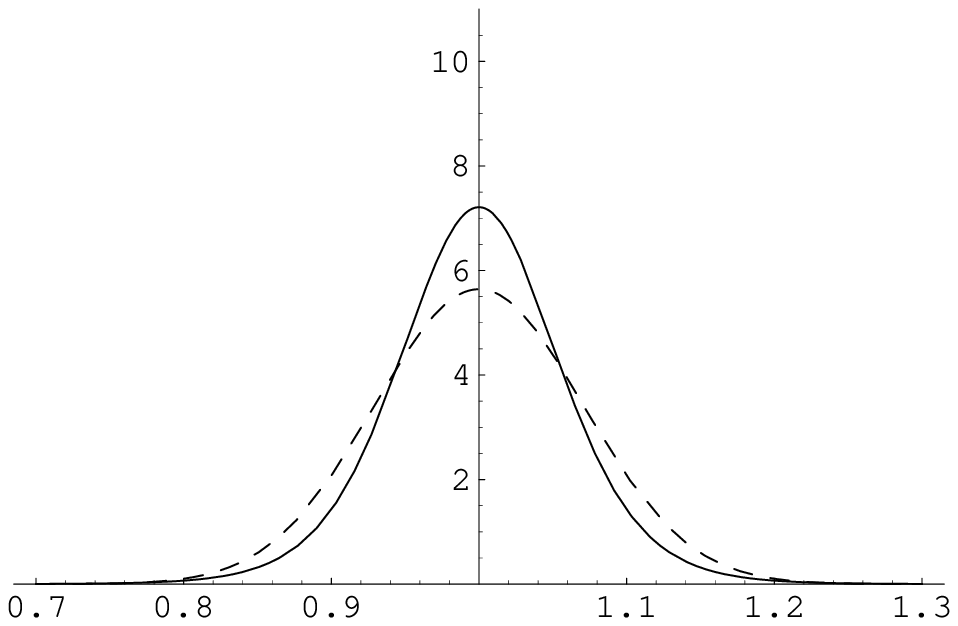,width=0.4\linewidth,clip=}  &
\epsfig{file=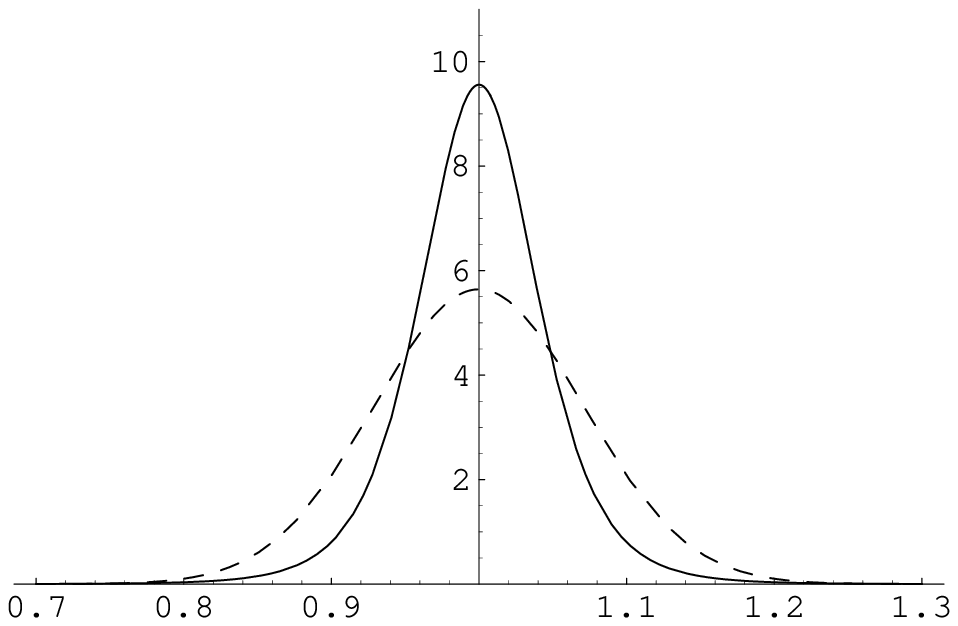,width=0.4\linewidth,clip=}  \\\hline
\epsfig{file=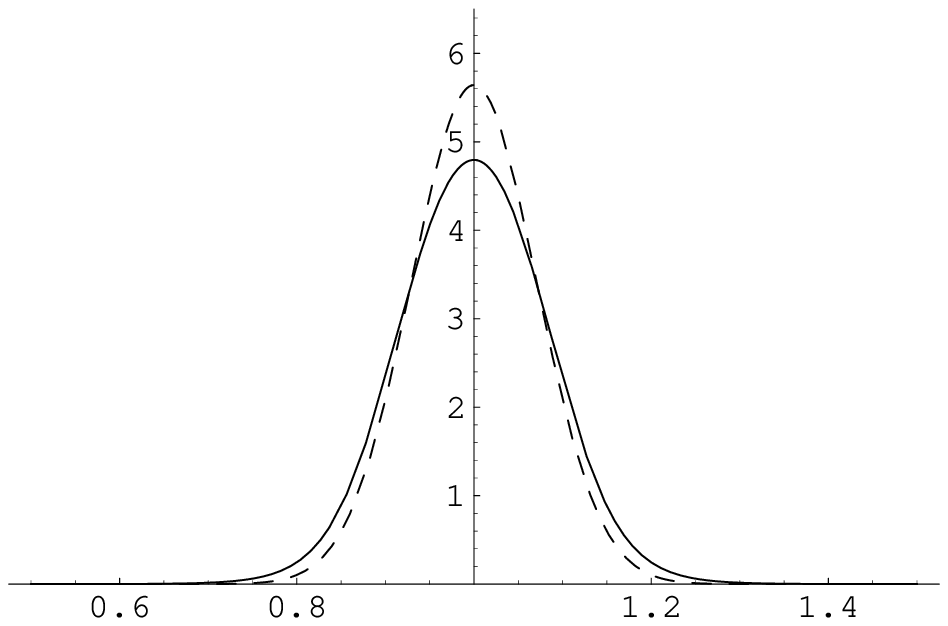,width=0.4\linewidth,clip=}  &
\epsfig{file=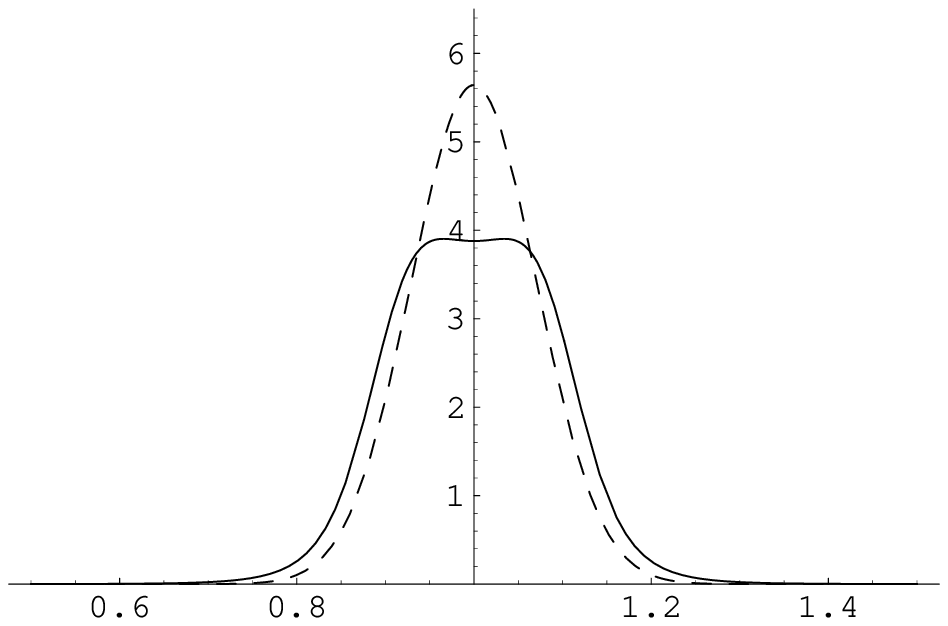,width=0.4\linewidth,clip=}  \\\hline
\epsfig{file=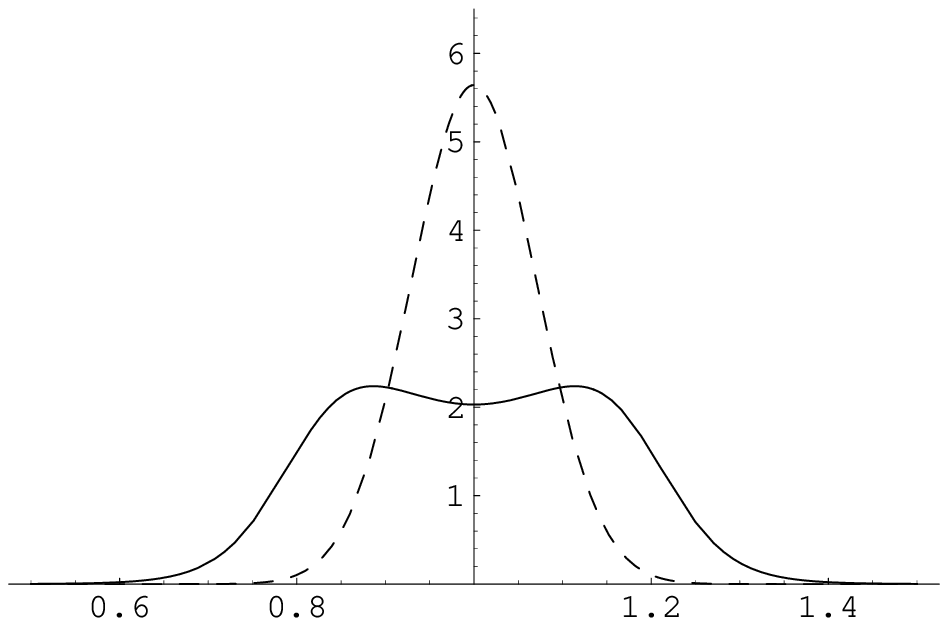,width=0.4\linewidth,clip=}  &
\epsfig{file=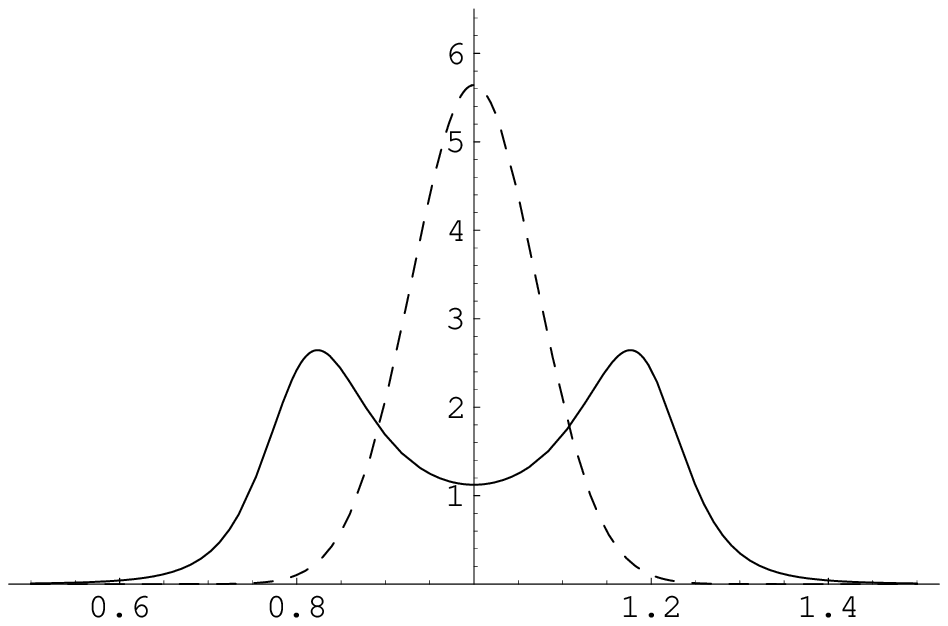,width=0.4\linewidth,clip=}  \\\hline
\epsfig{file=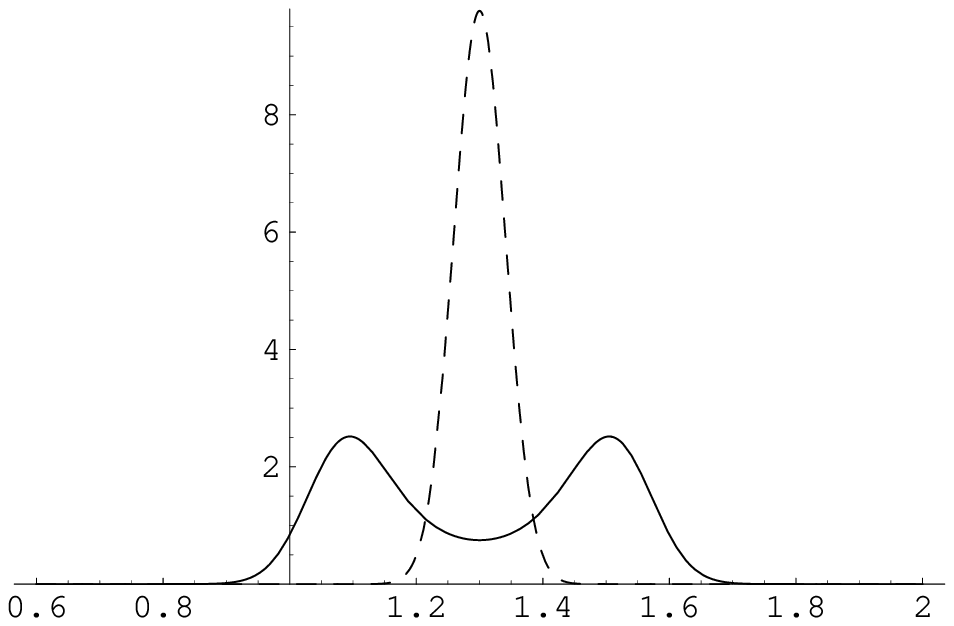,width=0.4\linewidth,clip=}  &
\epsfig{file=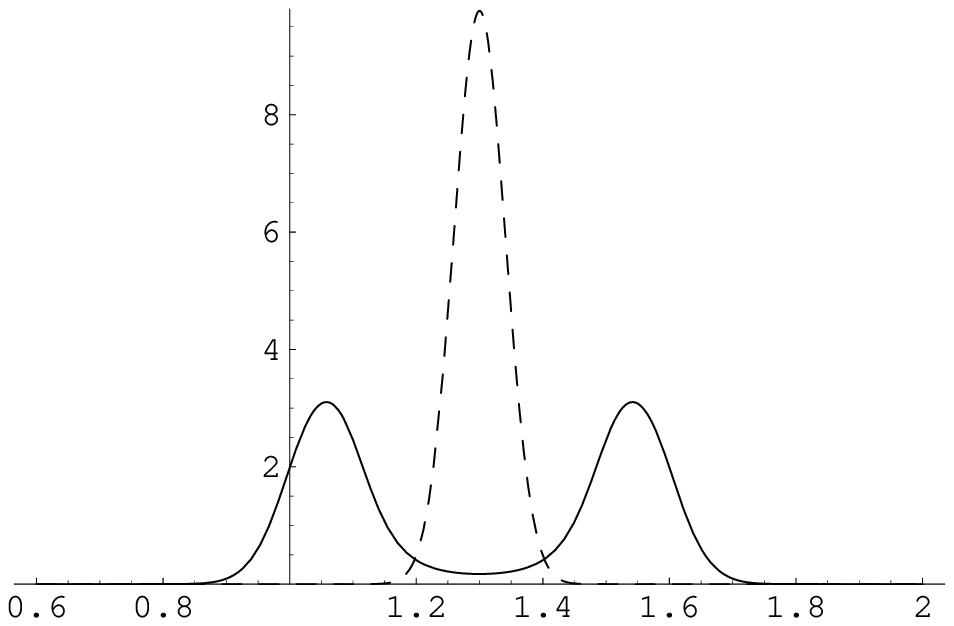,width=0.4\linewidth,clip=} \\ \hline
\epsfig{file=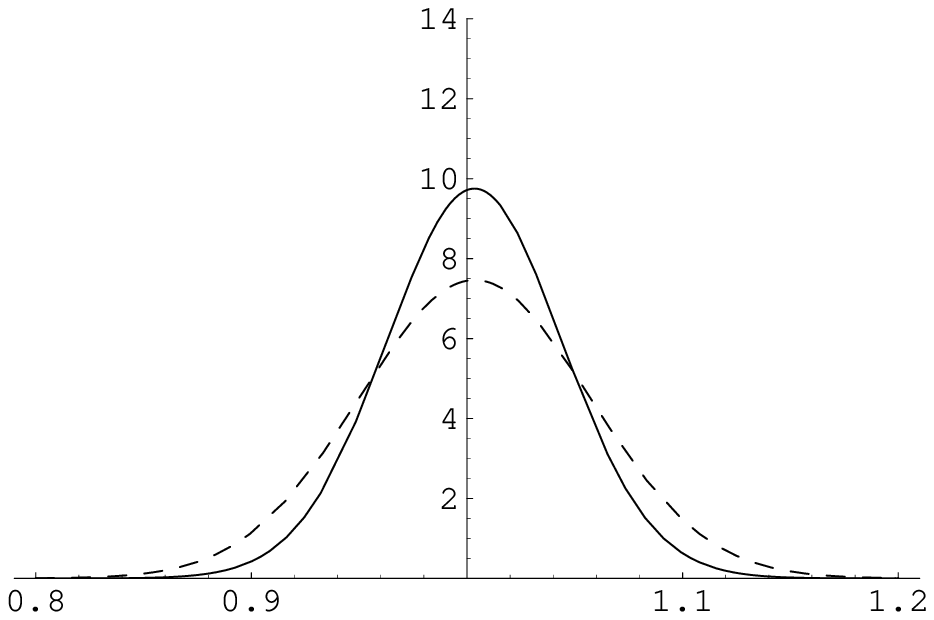,width=0.4\linewidth,clip=}  &
\epsfig{file=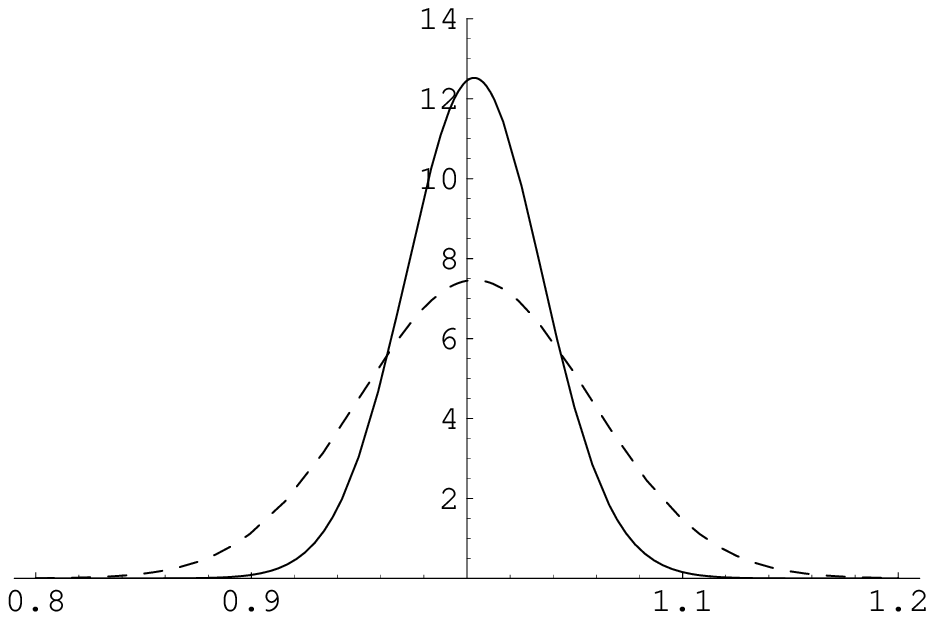,width=0.4\linewidth,clip=} \\ \hline
\end{tabular}
\end{center}
\caption{\small Combination of results obtained by 
varying the parameters of the sceptical combination.}
\label{fig:discordant3}
\end{figure}

\begin{figure}
\begin{center}
\begin{tabular}{|c|}\hline
\epsfig{file=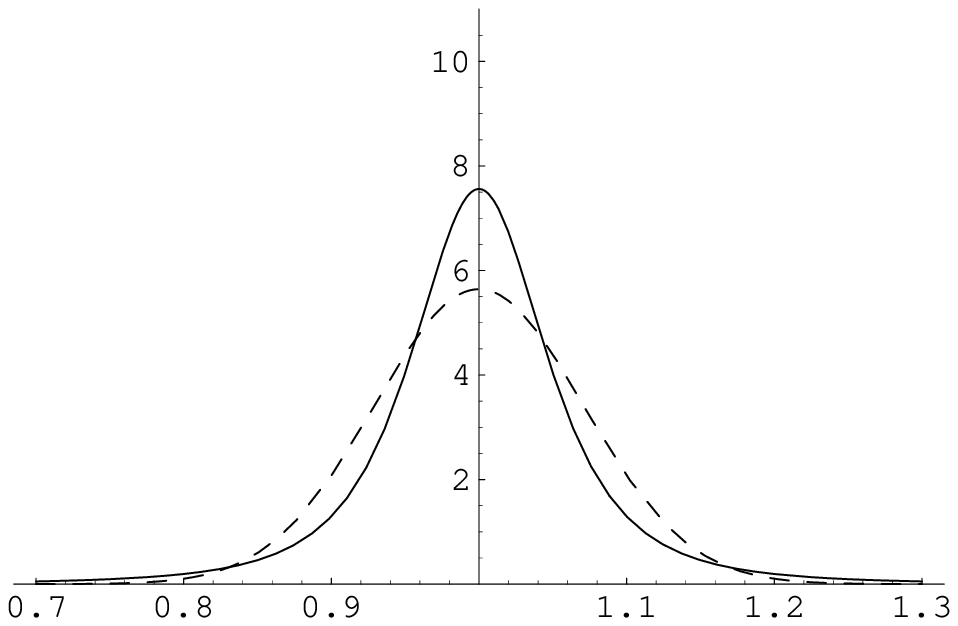,width=0.5\linewidth,clip=}  
\\ \hline
\end{tabular}
\end{center}
\caption{\small Sceptical perception of a single measurement 
having a standard deviation equivalent to the 
standard combination of the top of Fig.~\ref{fig:discordant}.
Note how the result differs from the combination of the individual results.}
\label{fig:single}
\end{figure}
 
For simplicity, all individual 
results are taken to have the same standard deviation 
(note that the upper left plot of Fig.~\ref{fig:discordant}
shows the situation of two identical results). 
The solid curve of the right-hand plots shows the combined result 
obtained using Eq.~(\ref{eq:final_DL2}) with $\lambda=0.6$ 
and $\delta=1.3$, yielding $\mbox{E}[r]=\sigma(r)=1$. 
For comparison, 
the dashed lines show also the result obtained by 
the standard combination. The method described in this paper, 
with parameters chosen by general considerations, 
tends to behave in qualitative agreement 
with the expected point of view of a sceptical experienced physicist. 
As soon as the individual results start to disagree, 
the combined distribution gets broader than the standard combination, 
and might become multi-modal if the results cluster in 
several places. However, if the agreement is somehow `too good'
(first and last case of Fig.~\ref{fig:discordant}) the 
combined distribution becomes narrower than the standard result. 

In order to get a feeling about the sensitivity of 
the results from the choice of the parameters, 
two other sets of parameters have been tried, keeping the 
requirement $\mbox{E}[r]=1$, but  
varying $\sigma(r)$ by $\pm$50\,\%: $\sigma(r)$ = 0.5 is 
obtained for $\lambda \approx$
1.4 and $\delta \approx$ 2.1; $\sigma(r) = 1.5$ is 
obtained for $\lambda \approx$ 0.4 and
$\delta \approx$ 1.1. The resulting p.d.f.'s of $r$ are shown 
in Fig. \ref{fig:rDL2}. The results obtained using these
two sets of parameters on the simulated data of 
Fig.~\ref{fig:discordant} are shown in Fig.~\ref{fig:discordant3}. We 
see that, indeed, the choice  
$\mbox{E}[r]=\sigma(r)=1$ seems to be an optimum, 
and the $\pm 50\%$ variations of $\sigma(r)$ give results 
which are at the edge of what one would consider to be 
acceptable. Therefore, we shall take the parameters 
providing $\mbox{E}[r]=\sigma(r)=1$ as the reference ones.

Another interesting feature of Eq.~(\ref{eq:final_DL2}) is its behaviour
for a single experimental result, as shown in Fig.~\ref{fig:single}. 
For comparison, we have taken a result  
 having a stated 
standard deviation equal to $1/\sqrt{2}$ of each of those of 
Fig.~\ref{fig:discordant}.  Figure \ref{fig:single} 
 has to be compared with the upper right
plots of Fig.~\ref{fig:discordant}. 
 The sceptical combination
takes much more seriously two independent  experiments, each
reporting in an uncertainty $\sigma$,  
than a single experiment performing $\sigma/\sqrt{2}$. 
On the contrary, 
the two situations are absolutely equivalent in the standard 
combination rule. In particular, the tails of the p.d.f. 
obtained by the sceptical
combination vanish  more slowly than in the Gaussian case, 
while the belief in the 
central value is higher. The result models 
 the qualitative attitude of sceptical 
physicists, according to whom a single experiment is 
never enough to establish a value, no matter how
precise the result may be, although the true value
might have more 
chance to be within one standard deviation than the
probability level 
calculated from a Gaussian distribution. 

\section{Application to $\epsilon^\prime/\epsilon$}
\begin{figure}[!t]
\begin{center}
\begin{tabular}{|c|}\hline
\epsfig{file=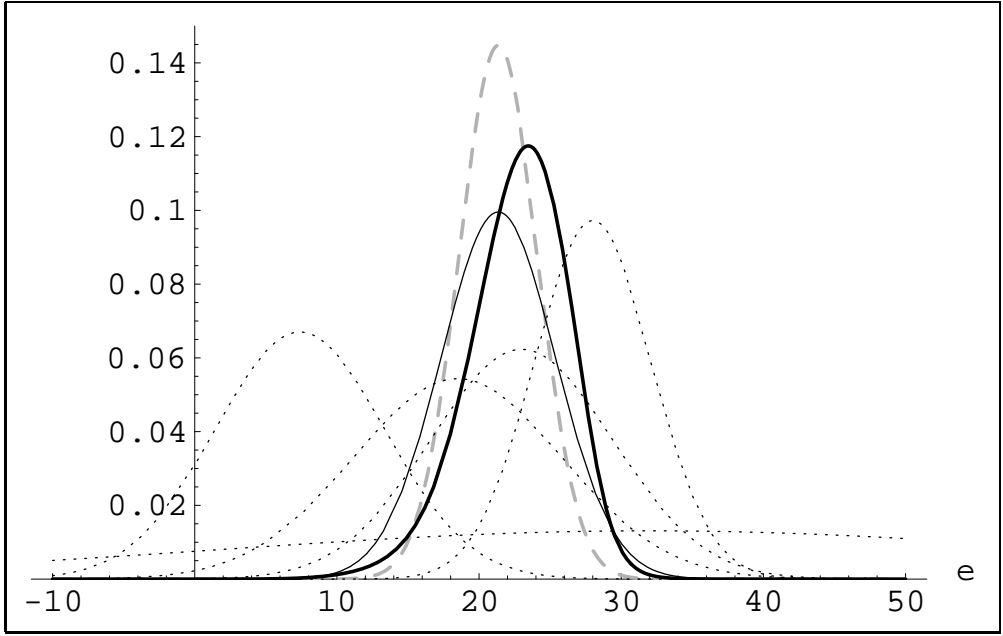,width=0.8\linewidth,clip=}
\\ \hline
\end{tabular} 
\end{center}
\caption{\small Individual results 
compared with the standard combination
(grey dashed), the PDG-rescaled combination (solid thin) and the
sceptical combination as described in this paper (solid thick).}
\label{fig:combinations}
\end{figure}

The combination rule based on Eq.~(\ref{eq:final_DL2}) 
has been applied to the results about Re($\epsilon^\prime/\epsilon$) 
shown in Table~\ref{tab:results}. As discussed above, 
our reference parameters are $\lambda=0.6$ and $\delta=1.3$,
corresponding to $\mbox{E}[r]\approx \sigma(r)\approx 1$. 
The resulting p.d.f. for $e$ = Re($\epsilon^\prime/\epsilon) \times 10^4$
 is shown as the thick continuous line of 
Fig.~\ref{fig:combinations}, together with the
individual results (dotted lines).  
For comparison, we also give
the result obtained using the combination rules commonly applied
in particle physics. The grey-dashed line 
of Fig.~\ref{fig:combinations} is obtained 
with the standard combination rule [Eqs. (\ref{eq:media})
and (\ref{eq:sigma})]. The thin continuous line 
has been evaluated using the Particle Data Group (PDG) 
`prescription'~\cite{PDG98}.
According to this rule, the standard deviation  
(\ref{eq:sigma}) is enlarged by a factor given by 
$\sqrt{\chi^2/(N-1)}$, where $\chi^2$ is the chi-2 of the data
with respect to the average (\ref{eq:media}) and $N$ is
the number of independent results.\footnote{Note that the `official' 
world average obtained
using the PDG recipe of $(21.2 \pm 0.46 \times  10^{-4}$ (see e.g. 
\cite{Buras,Fabbrichesi,Roma99})  differs
from that given here because all five  results of Table 1 are used here,  
as I do not see
any reason why the 1988 E731 result should be disregarded.}
\begin{table}[!b]
\caption{\small Comparison of the 
different methods of combining the results.}
\label{tab:comparison}
\begin{center}
\begin{tabular}{|c|c|c|c|c|c|}\hline
&&&&& \\
Combination & Mean ($\sigma$) & Median & Mode  & 99\% range& 
$P[{\rm Re}(\epsilon^\prime/\epsilon<0)]$ \\ 
&&& $\pm\,34\%$ range   && \\ \hline 
&&&&& \\
Standard    & $21.4\, (2.7)$& 21.4  & $21.4\pm2.7$ & $[14.3, 28.5]$ & 
$5\times 10^{-15}$ \\
&&&&& \\
PDG rule~\cite{PDG98} & $21.4\, (4.0)$ &21.4  & $21.4\pm4.0$ & 
$[11.0, 31.7]$ & 
$5\times 10^{-8}$ \\
&&&&& \\
Sceptical &  22.7 (3.5) &  23.0 &  
 $23.5\pm3.4$ & $[11.6, 30.5]$ & $1.5\times 10^{-6}$ \\ 
&&&&& \\
\hline
\end{tabular}
\end{center}
\end{table}

We see that
although  the PDG rule gives a distribution wider than 
that obtained  by the standard
rule, the barycentres of the distributions coincide, 
thus not taking into account that one
of the  results is quite far from 
where the others seem to cluster.  Moreover, the p.d.f.
is assumed to be Gaussian, independently  
of the configuration of experimental points.
 Instead, the sceptical combination takes 
into account better the configuration of the data points. 
The peak of the distribution is essentially determined
by the three results which appear more consistent with each 
other. Nevertheless, there is a more pronounced tail 
for small values of Re($\epsilon^\prime/\epsilon$), to take 
into account that there is indeed a result providing evidence 
in that region, and that cannot be ignored. 

A quantitative comparison of the different methods is
given in Table~\ref{tab:comparison}, 
where the most relevant statistical summaries are provided 
(average, mode, median, standard deviation), together 
with some probability intervals. It is worth recalling 
that each of  these summaries gives some information about the distribution, 
but, when the  uncertainty
of this result has to be finally 
propagated into other results  
(together with other uncertainties), 
it is the  average and standard deviation
which matter.\footnote{The standard 
`error propagation' is based on linearization, on the property 
of expected value and variance under a linear combination 
and on central limit theory (the result of several contributions
will be roughly Gaussian). Therefore, propagating mode (or median) 
and 68\% probability intervals does not make any sense, 
unless the input distributions are Gaussian.} 
An interesting comparison is given by the probability that 
Re($\epsilon^\prime/\epsilon$) is negative. The sceptical combination 
gives the largest value, but still at the level of one part per million, 
indicating that, even in this conservative analysis, 
a positive value of the direct CP violation parameter 
seems `practically' established. 

\begin{figure}[t]
\begin{center}
\begin{tabular}{|c|}\hline
\epsfig{file=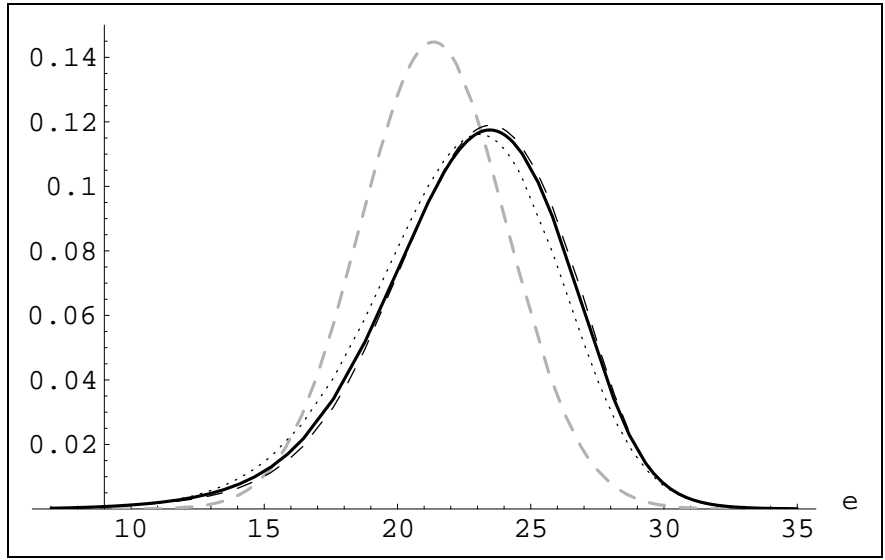,width=0.7\linewidth,clip=} 
\\ \hline
\end{tabular}
\end{center}
\caption{\small Dependence of the sceptical combination 
on the choice of the parameters. Continuous, dotted and 
dashed lines are, in order:   
$\lambda=0.6$ and $\delta=1.3$ $[\sigma(r)=1)]$;   
$\lambda=0.4$ and $\delta=1.1$ $[\sigma(r)=0.5)]$;
$\lambda=1.4$ and $\delta=2.1$ $[\sigma(r)=0.5)]$. 
The grey-dashed line gives, for comparison, the result of
the standard combination.}
\label{fig:var_result}
\end{figure}

The sensitivity of the result on the parameters of the 
combination formula can be inferred from Fig.~\ref{fig:var_result}, 
where the results obtained changing $\sigma(r)$ by $\pm 50\%$
are shown. The combined result is quite stable. This is
particularly true if one remembers that these extreme 
values of parameters are quite at the edge of what 
one would accept as reasonable, as   can be seen in 
Fig.~\ref{fig:discordant3}. Note that if one would like
to combine the results taking also into account the uncertainty 
about the parameters, one would apply Eq.~(\ref{eq:intld}). 
It is reasonable to think that, 
since the variations of the p.d.f. from that obtained 
for the reference  value of the parameters are not very large,  
the p.d.f. obtained as weighted average over all the possibilities
will not be much different from the reference one. 

\begin{figure}[t]
\begin{center}
\begin{tabular}{|c|}\hline
\epsfig{file=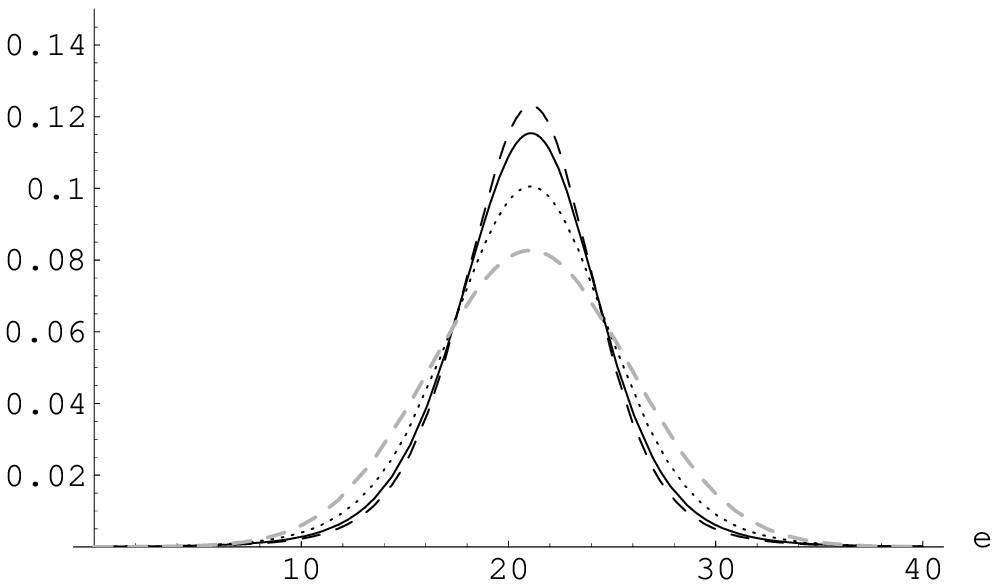,width=0.65\linewidth,clip=}  \\ 
\hline
\epsfig{file=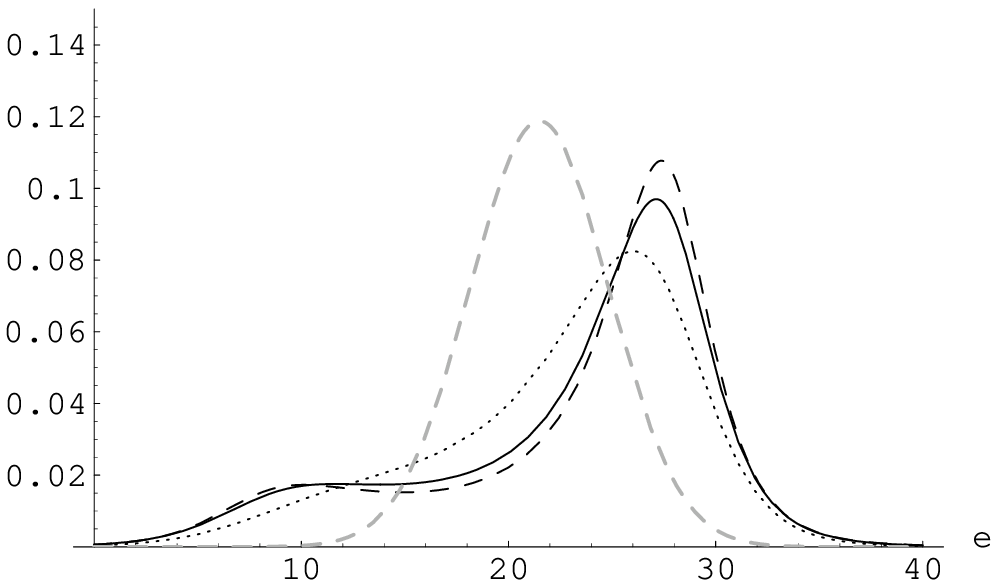,width=0.65\linewidth,clip=} \\ \hline
\end{tabular}
\end{center}
\caption{\small Sceptical combination of CERN and Fermilab results
(upper and lower plot, respectively). The continuous line 
shows the result obtained by Eq.~(\ref{eq:final_DL2}) and 
reference parameters. The dashed and dotted lines are the results
obtained by varying the standard deviation of $r=\sigma/s$ by 
$+50\%$ and $-50\%$, respectively. The grey-dashed line shows 
the results obtained by the standard combination rule.} 
\label{fig:partialDL}
\end{figure}

 Figure~\ref{fig:partialDL} and Table~\ref{tab:comparison1}
give the results subdivided into CERN and Fermilab.
 In these cases the difference between the standard combination 
and the sceptical combination becomes larger, and, again,  
the outcome of the sceptical combination follows qualitatively 
the intuitive one of
experienced physicists. The sceptical combination of the CERN results 
alone is better than that given by the standard one, thus
reproducing formally the
\break\hfill
\begin{table}[H]
\caption{\small Comparison of the different methods of combining partial
 results. The symbol * means that the distribution has less 
than 34.1\% probability on the right side of the mode.}
\label{tab:comparison1}
\begin{center}
{\small{\begin{tabular}{|cr|c|c|c|c|c|}\hline
\multicolumn{2}{|c|}{} &&&&& \\
\multicolumn{2}{|c|}{Combination} 
& Mean ($\sigma$) & Median & Mode & 99\% p. range& 
$P[\mbox{Re}(\epsilon^\prime/\epsilon)<0]$ \\ 
\multicolumn{2}{|c|}{} &&&$\pm 34\%$ p. range && \\ \hline 
&&&&&& \\
Stand. & $\left\{\begin{array}{l} \mbox{CERN} \\ 
                               \mbox{Fermilab} \end{array}\right.$& 
$\begin{array}{l} 21.1\, (4.8) \\ 21.4\, (2.7) \end{array}$ &
$\begin{array}{l} 21.1   \\ 21.4                \end{array}$ &
$\begin{array}{l} 21.1 \pm 4.8   \\  21.4\pm2.7  
        \end{array}$ &
$\begin{array}{c}
  \mbox{[8.6, 33.4]} \\  \mbox{[12.9, 30.1]}   
        \end{array}$ &
$\begin{array}{l} 6\times 10^{-6} \\ 
                      8\times 10^{-11} \end{array}$ \\
&&&&&& \\
Scept. & $\left\{\begin{array}{l} \mbox{CERN} \\ 
                               \mbox{Fermilab} \end{array}\right.$ & 
$\begin{array}{l} 21.0\, (3.9) \\ 23.0\, (7.1) \end{array}$ &
$\begin{array}{l} 21.0   \\ 25.2               \end{array}$ &
$\begin{array}{l} 21.1 \pm 3.6   \\  27.1^{+\,*}_{- 4.9}  
        \end{array}$ &
$\begin{array}{c}   \mbox{[9.2, 32.5]} \\  
                    \mbox{[2.7, 36.2]}
        \end{array}$ &
$\begin{array}{l} 2.5\times 10^{-4} \\ 
                      1.5\times 10^{-3} \end{array}$ \\
&&&&&& \\
\hline
\end{tabular}}}
\end{center}
\end{table}
\newpage\noindent
instinctive suspicion that the 
uncertainties could have been overestimated. 
For the Fermilab ones the situation is reversed. 
In any case, both partial combinations tend to establish 
strongly the picture
of a positive and sizeable Re($\epsilon^\prime/\epsilon$) value.
Finally, note that the $\pm 50$\% variations in $\sigma(r)$ produce 
in the partial combinations a larger effect (although not 
relevant for the conclusions) than in the 
overall combination. This is due to the fact that the 
variations produce opposite effects on the two subsets of data in 
the region of Re($\epsilon^\prime/\epsilon$) around $20\times 10^{-4}$.

\section{Posterior evaluation of $\sigma_i$}
An interesting by-product of the method illustrated above is 
the posterior evaluation of the various $\sigma_i$, or, equivalently, 
of the various $r_i$. Again, we can make use of Bayes' theorem, 
obtaining
\begin{equation}
f(\underline{r}\,|\,\underline{d},\underline{s},\mu) =
\frac{f(\underline{d}\,|\,\underline{r},\underline{s},\mu)
\cdot f_\circ(\underline{r}\,|\,\underline{s},\mu)}
{\int\!f(\underline{d}\,|\,\underline{r},\underline{s},\mu)
\cdot f_\circ(\underline{r}\,|\,\underline{s},\mu)
\,\mbox{d}\underline{r}}~,
\label{eq:Bayes_r}
\end{equation}
\noindent
where $\underline{r}=\{r_1, r_2,\ldots, r_n\}$. 
Since the initial 
status of knowledge is such that values of  $r_i$ are    
independent of each other, and they are independent of 
$\mu$ and $\underline{s}$, we obtain 
\begin{equation}
f_\circ(\underline{r}\,|\,\underline{s},\mu) = 
f_\circ(\underline{r}) = \prod_if_\circ(r_i) \equiv 
\prod_i f(r_i\,|\,\lambda,\delta) = 
\prod_i \frac{2\,\lambda^\delta\,
r_i^{-(2\,\delta+1)}\,e^{-\lambda/r_i^2}}{\Gamma(\delta)}~,
\label{eq:f0r}
\end{equation}
having used Eq.~(\ref{eq:r_Dose1}). As a shorthand for Eq.~(\ref{eq:f0r}), 
we shall write in the following
simply $f_\circ(\underline{r})=\prod_if_\circ(r_i)$.

Since  the experimental results are also considered independent,
we can rewrite Eq.~(\ref{eq:Bayes_r}) as 
\begin{eqnarray}
f(\underline{r}\,|\,\underline{d},\underline{s},\mu) &=&
\frac{\prod_i f(d_i\,|\,r_i,s_i,\mu)
\cdot f_\circ(r_i)}
{\int\!\prod_i f(d_i\,|\,r_i,s_i,\mu)
\cdot f_\circ(r_i)\, \mbox{d}\underline{r}} \nonumber \\
&= &\frac{\prod_i f(d_i\,|\,r_i,s_i,\mu)
\cdot f_\circ(r_i)}
{\prod_i \int\!f(d_i\,|\,r_i,s_i,\mu)
\cdot f_\circ(r_i)\, \mbox{d}r_i}\,~. 
\label{eq:Bayes_r2}
\end{eqnarray}
The marginal distribution of each $r_i$, still conditioned by $\mu$
(and, obviously, by the experimental values), 
is  obtained  by integrating 
$f(\underline{r}\,|\,\underline{d},\underline{s},\mu)$ over 
all $r_j$, with $j\ne i$. As a result, we obtain
\begin{eqnarray}
f(r_i\,|\,\underline{d},\underline{s},\mu) &=&
\frac{f(d_i\,|\,r_i,s_i,\mu)
\cdot f_\circ(r_i)}
{\int\! f(d_i\,|\,r_i,s_i,\mu)
\cdot f_\circ(r_i)\, \mbox{d}r_i}\,~.
\label{eq:Bayes_r3}
\end{eqnarray}
Making use of Eqs. (\ref{eq:lik_i_r}), (\ref{eq:r_Dose1}) and 
(\ref{eq:int_lik2}) we get: 
\begin{eqnarray}
f(r_i\,|\,\underline{d},\underline{s},\mu) &=&
\frac{\frac{1}{\sqrt{2\,\pi}\,r_i\,s_i}\,
                     \exp\left[-\frac{(d_i-\mu)^2}{2\,r_i^2\,s_i^2}\right]
      \cdot 
      \frac{2\,\lambda^\delta\,
      r_i^{-(2\,\delta+1)}\,e^{-\lambda/r_i^2}}{\Gamma(\delta)}}
     {\frac{\lambda^\delta}{\sqrt{2\,\pi}s_i}\,
                 \frac{\Gamma(\delta+1/2)}{\Gamma(\delta)}\,
       \left(\lambda+\frac{(d_i-\mu)^2}
                          {2\,s_i^2}\right)^{-(\delta+1/2)}}\,.
\label{eq:f_r_i_cond_mu}
\end{eqnarray}
The final result is obtained by eliminating, in the usual way, the 
condition $\mu$, i.e. 
\begin{equation}
f(r_i\,|\,\underline{d},\underline{s}) =
\int\! f(r_i\,|\,\underline{d},\underline{s},\mu) \cdot
f(\mu\,|\,\underline{d},\underline{s}) \,\mbox{d}\mu\,~.
\label{eq:f_r_i}
\end{equation}
Making use of Eq.~(\ref{eq:final_DL2}), and 
neglecting in Eq.~(\ref{eq:f_r_i_cond_mu})
all factors not depending on $r_i$ and $\mu$, we 
get the unnormalized result
\begin{equation}
f(r_i\,|\,\underline{d},\underline{s}) 
\propto 
r_i^{-(2\,\delta+2)}\,e^{-\lambda/r_i^2}
\int\!\exp\left[-\frac{(d_i-\mu)^2}{2\,r_i^2\,s_i^2}\right]
\,\prod_{j\ne i}
\left(\lambda+\frac{(d_j-\mu)^2}{2\,s_j^2}\right)^{-(\delta+1/2)}
\,\mbox{d}\mu\,~.
\label{eq:unn_r_i}
\end{equation} 
This formula is clearly valid for $n\ge 2$. 
If this is not the case, the product over $j\ne
i$ is replaced by unity,  and the integral is proportional to $r_i$.
Equation~(\ref{eq:unn_r_i}) becomes then
$f(r_1\,|\,d_1,s_1) \propto r_1^{-(2\,\delta+1)}\,e^{-\lambda/r_1^2}$,
i.e. we have recovered the initial distribution (\ref{eq:r_Dose1}). 
In fact, if we have only one data point, there is no reason to
change our beliefs about $r$. Only the comparison with other results 
can induce us to change our opinion. 

Once we have got $f(r_i\,|\,\underline{d},\underline{s})$ we can give
posterior estimates of $r_i$ in terms of average and standard deviations, 
and they can be compared with the prior assumption 
$\mbox{E}[r]=\sigma(r)=1$, to understand which uncertainties have been 
implicitly rescaled by the sceptical 
combination.\footnote{Note that it is incorrect to feed again into the
procedure the rescaled uncertainties, as they come  from this analysis. 
The procedure has  already taken into account   all possible 
rescaling factors in the evaluation of 
$f(\mu\,|\,\underline{d},\underline{s})$.}  
 Convenient formulae
to evaluate numerically first and second moments of the 
posterior distribution of $r_i$ are given by\footnote{Note that, since 
$\prod_j(\ldots)$ of the integrands are proportional 
to $f(\mu\,|\,\underline{d},\underline{s})$, 
Eqs.~(25)--(26)
can be written in the compact form
\begin{eqnarray*}
\mbox{E}[r_i] &=& \frac{\Gamma(\delta)}{\Gamma(\delta+1/2)}
\cdot \mbox{E}_{\mu}\left[
\left(\lambda+\frac{(d_i-\mu)^2}{2\,s_i^2}\right)^{1/2}\right] \\
\mbox{E}[r_i^2] &=& \frac{\Gamma(\delta-1/2)}{\Gamma(\delta+1/2)}
 \cdot \mbox{E}_{\mu}\left[             
 \lambda+\frac{(d_i-\mu)^2}{2\,s_i^2} \right]\,,
\end{eqnarray*}
where $\mbox{E}_\mu[\cdot]$ indicates expected values over 
the p.d.f. of $\mu$.}
\begin{eqnarray}
\mbox{E}[r_i] &=& \frac{\Gamma(\delta)}{\Gamma(\delta+1/2)}
\cdot \frac{\int\!\left(\lambda+\frac{(d_i-\mu)^2}{2\,s_i^2}\right)^{1/2}
            \prod_j\left(\lambda+\frac{(d_j-\mu)^2}{2\,s_j^2}
                   \right)^{-(\delta+1/2)}\,\mbox{d}\mu}
           {\int\! \prod_j\left(\lambda+\frac{(d_j-\mu)^2}{2\,s_j^2}
                   \right)^{-(\delta+1/2)}\,\mbox{d}\mu} \\
\mbox{E}[r_i^2] &=& \frac{\Gamma(\delta-1/2)}{\Gamma(\delta+1/2)}
\cdot \frac{\int\!\left(\lambda+\frac{(d_i-\mu)^2}{2\,s_i^2}\right)
            \prod_j\left(\lambda+\frac{(d_j-\mu)^2}{2\,s_j^2}
                   \right)^{-(\delta+1/2)}\,\mbox{d}\mu}
           {\int\! \prod_j\left(\lambda+\frac{(d_j-\mu)^2}{2\,s_j^2}
                   \right)^{-(\delta+1/2)}\,\mbox{d}\mu}\,~.
\end{eqnarray}
At this point it is important to anticipate the objection
of those who think that it is incorrect to infer 
$n+1$ quantities ($\mu$ and $\underline{r}$) 
starting from $n$ data points. 
Indeed, there is nothing wrong in doing so. But, obviously, 
the results are correlated, and they depend also on the prior 
distribution of $r_i$, which acts as a constraint. In fact
we have seen above that for $n=1$ the result on $r$ is trivial. 

\begin{figure}[t]
\begin{center}
\begin{tabular}{|c|}\hline
\epsfig{file=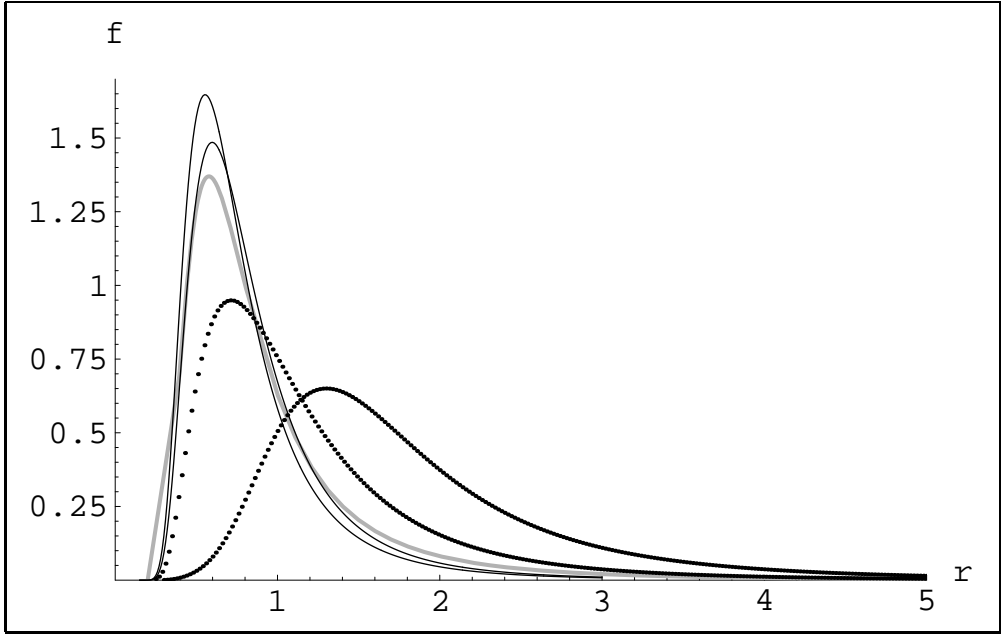,width=0.8\linewidth,clip=}  \\ \hline
\end{tabular}
\end{center}
\caption{\small Final distributions of $r$ corresponding
to the four most precise results on Re($\epsilon^\prime/\epsilon$), 
compared with the 
reference prior one (grey line), i.e. having $\mbox{E}[r_i]=\sigma(r_i)=1$
$\forall i$. The continuous lines refer to the CERN results, 
dotted lines to the Fermilab ones.}
\label{fig:final_r}
\end{figure}

Figure \ref{fig:final_r} gives the final distributions of 
$r_i=\sigma_i/s_i$ for the four most precise 
determinations of Re($\epsilon^\prime/\epsilon$) 
(the 1988 E731 result has not been
plotted because it is very similar to the NA31 one, as one
can understand from Table~\ref{tab:posterior_r}), 
compared with the reference 
initial distribution having $\sigma(r)=1$ (grey line in the plot).
The distributions relative to the CERN results are shown with 
continuous lines, the Fermilab ones by dots. In particular, 
the one that has a
substantial probability mass above 1 is the 1993 E731 result.  
Average\break\hfill\newpage\noindent 
and standard
deviations of the distributions are  given in Table~\ref{tab:posterior_r}, 
also showing the values
 that one would obtain with the other sets of parameters
that we have considered to be edge ones.

Once more, the results are in qualitative agreement with 
intuition: The CERN curves are slightly squeezed below $r=1$, 
as the uncertainty evaluation seems to be a bit conservative.  
The Fermilab ones show instead some drift towards large $r$. 
In particular,
figure and table make one suspect that 
some contribution to the
error has been overlooked in the E731 data. Note that in this case 
the average value of the rescaling factor 
is smaller than one could 
expect from alternative procedures which require 
the overall $\chi^2$ to equal the number of degrees of freedom. 
The reason is the shape of the  
initial distribution of $r$, which  protects us against 
unexpectedly large values of the rescaling factors. 
\begin{table}[H]
\caption{\small Posterior estimation of $r=\sigma_i/s_i$ 
starting from identical priors having $\mbox{E}_\circ[r]=1$ 
and $\sigma_\circ(r)=0.5$, 1.0 and 1.5. The individual 
results are given by $d_i\pm s_i$ 
to be consistent with the notation used
throughout this paper.}
\label{tab:posterior_r}
\begin{center}
\begin{tabular}{|l|cc|ccc|}\hline
&&&&&\\
\multicolumn{1}{|c|}{Experiment} & $d_i$  &  $s_i$   & 
\multicolumn{3}{|c|}{Posterior $\mbox{E}[r_i]$ $(\sigma(r_i))$} \\
&&& $\sigma_\circ(r)=0.5$ & $\mathbf{\sigma_\circ(r)=1} $ &
   $\sigma_\circ(r)=1.5$ \\
&&&&&\\
\hline
E731 (1988) \cite{E73188}          & 32  & 30     & 
0.9 (0.4) & 0.8  (0.5) & 0.7 (0.5) \\
E731 (1993)\cite{E73193}           & 7.4  & 5.9   &
1.6 (0.7) &  1.9  (1.2)& 2.1 (1.5) \\
NA31 (1988+1993)\cite{NA3193,Wahl} &23.0  & 6.5   & 
0.9 (0.4) & 0.8  (0.5) & 0.8 (0.6) \\
KTeV (1999)\cite{KTeV}             & 28.0  & 4.1  &
1.2 (0.6) &  1.2  (0.9)& 1.2 (1.0) \\
NA48 (1999)\cite{NA48}             & 18.5  & 7.3  & 
0.9 (0.4) &  0.9  (0.5)& 0.9 (0.6) \\
\hline
\end{tabular}
\end{center}
\end{table}

\section{Discussion and conclusions}
The problem of combining data which appear in mutual disagreement 
has been analysed from a probabilistic perspective. 
We have started from the usual hypotheses on which the 
well-known combination rule is based and  
we have seen that a possible solution can be based 
on a suitable modelling of the uncertainty on the 
standard deviation which describes the Gaussian likelihood. 
The complete status of uncertainty on the true value 
resulting from the various pieces of information 
is quantified by a p.d.f. $f(\mu)$ which, in our approach, 
does not have
an a priori defined shape. This property allows one to obtain 
results which never conflict with the intuitive judgement
of experienced physicists. The method described here 
 also allows one to infer the ratio between the `true' standard 
deviation and the stated one, as a result of the mutual agreement 
of the data. 

The application of this method to CP violation results 
from K$^0\rightarrow 2\pi$ shows that 
 the picture of a positive and sizeable value 
of Re($\epsilon^\prime/\epsilon$) survives a sceptical analysis.
This conclusion  also holds if one considers separately 
CERN and Fermilab results. As far as a number to summarize 
the result is concerned, the mass of probability is 
concentrated around $23.5\times 10^{-4}$, with a 
$\pm 3.4\times 10^{-4}$ interval having a 68\%  probability
of  containing  the true value. However the p.d.f. has 
a negative skewness that cannot be ignored. As a consequence, 
the expected value is slightly below the mode, 
at  $22.7\times 10^{-4}$. We would like to re-state  
that what matters for uncertainty propagation is the expected 
value, 
together with the standard deviation ($3.5\times 10^{-4}$), 
and not the mode, or the median, and the 
$\pm34\%$ probability interval around either of them. 

\begin{figure}
\begin{center}
\epsfig{file=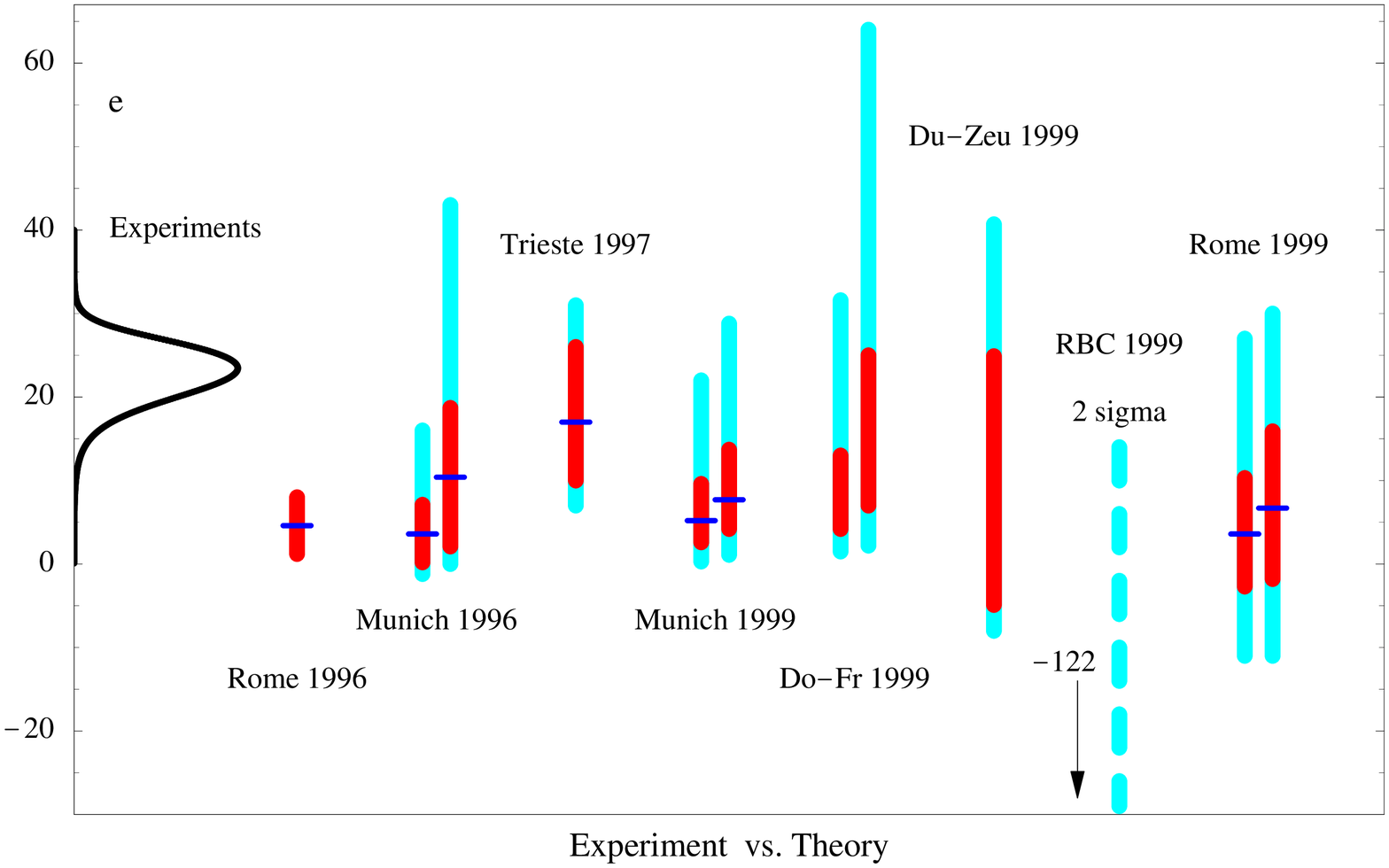,width=\linewidth,clip=}
\end{center}
\caption{\small Combined result on Re($\epsilon^\prime/\epsilon$)
compared with recent and very 
new theoretical calculations
(see text). }
\vspace{-2.3cm}\mbox{}
\hspace{32.4mm}\cite{Ciuchini}
\hspace{7.8mm}\cite{Monaco96}
\hspace{8.mm}\cite{Trieste97}
\hspace{7.8mm}\cite{Monaco99}
\hspace{8.0mm}\cite{Dortmunt99}
\hspace{8.0mm}\cite{Dubna99}
\hspace{11mm}\cite{RBC99}
\hspace{3.6mm}\cite{Roma99}
\vspace{+1.8cm}
\label{fig:fabbrichesi}
\end{figure}

\begin{figure}[t]
\begin{center}
\epsfig{file=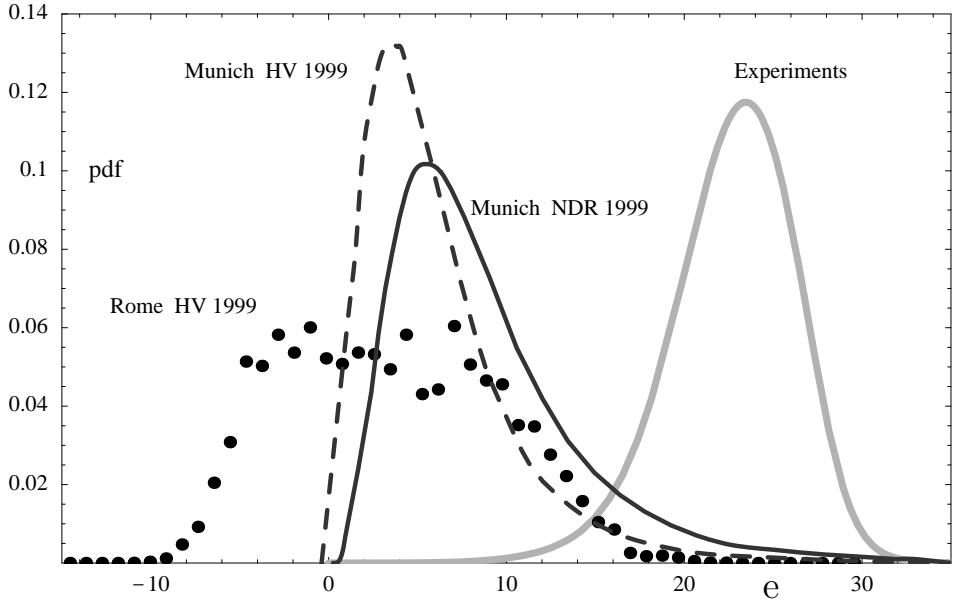,width=0.8\linewidth,clip=}
\end{center}
\caption{\small Probability density functions
 resulting from the combined experimental
information about Re($\epsilon^\prime/\epsilon$) 
compared with 1999 theoretical evaluations by the 
Munich~\protect{\cite{Monaco99}} and the Rome~\protect{\cite{Roma99}} groups
(the Rome NDR evaluation is very similar to the HV one).}
\label{fig:confr_pdf}
\vspace{-2.4cm}
\mbox{ }\hspace{11.5cm}
{\large e}
\vspace{1.7cm}
\end{figure}

The 1999 experimental results on 
Re($\epsilon^\prime/\epsilon$) have indeed renewed the 
interest of theorists in the subject. 
The comparison of the combined result 
with recent \cite{Ciuchini,Monaco96,Trieste97} 
and very new \cite{Monaco99,Dortmunt99,
Dubna99,RBC99,Roma99} theoretical 
evaluations~ is given in 
Fig.~\ref{fig:fabbrichesi}, an extension of the
updated version~\cite{Marco}
of Fig.~2 of Ref.~\cite{Fabbrichesi}. 
The vertical bands quantify somehow the 
uncertainty stated by the theoretical teams. 
The dark-grey bars should have the meaning of 68\% 
central probability bands, although sometimes they
are given as standard deviation of a non-Gaussian 
distribution. 
The grey bars are obtained using a procedure
that the theorists call `scanning' (see original papers),
but which has no well-defined probabilistic meaning. 
Since scanning produces very pessimistic uncertainty 
intervals, covering values of Re($\epsilon^\prime/\epsilon$)
which the authors hardly believe, one should be 
careful about concluding from Fig.~\ref{fig:fabbrichesi}
that the experimental value of  Re($\epsilon^\prime/\epsilon$)
is well compatible with all the approaches used to evaluate it. 
For example, Fig.~\ref{fig:confr_pdf},
which shows the p.d.f.'s
of the Munich and Rome teams, alongside 
that obtained from the combined analysis of the experimental
results, gives a better
idea of the mutual compatibility, and  
of how to interpret
the grey bars of Fig.~\ref{fig:fabbrichesi}
(note, in particular the positive
skewness of the theory curves and negative skewness
 of the experiment curve).
The grey-dashed bar shows the upper 2\,$\sigma$ tail
of the result 
of a recent evaluation~\cite{RBC99} 
which gives a very large negative value, 
having also a large uncertainty. 

In conclusion, it seems that, given the well-known
difficulties both in the
experimental determination and in the theoretical evaluation,
the overall picture is not dramatically worrying 
(and therefore invoking 
new phenomenology seems premature). What it is 
{\it practically} certain is that direct CP violation in the 
neutral-kaon system is established. We are all looking forward to 
an accurate theoretical explanation of the effect.

\section*{Acknowledgements}
I am indebted to Volker Dose for having suggested
to me the two-parameter extension of his work 
with Wolfgang von der Linden, for cross-checking 
some of the final formulae, and for comments on 
an initial version of the paper. I have benefited 
from discussions on the subject with  Pia Astone and 
Heinrich Wahl, who  also gave me helpful comments 
on the manuscript.
I wish to thank
Andrzej Buras, Marco Ciuchini,
Marco Fabbrichesi, Enrico Franco and Guido Martinelli  
for useful communications and discussions. 
%I have benefited from 
%several discussions on the subject with Pia Astone and 
%Heinrich Wahl, which also gave me helpful comments 
%on the manuscript.

\end{document}